\documentclass[
reprint,
superscriptaddress,
amsmath,amssymb,
aps,
prb,
]{revtex4-2}
\usepackage{graphicx}
\graphicspath{{./Figure}}
\usepackage{dcolumn} 
\usepackage{bm} 
\usepackage{braket}
\usepackage{xcolor}
\usepackage{dsfont}
\usepackage{wasysym}
\usepackage[colorlinks, allcolors=blue]{hyperref}
\usepackage{xspace}
\usepackage{makecell}
\usepackage[table]{xcolor}
\usepackage{blkarray}
\usepackage{multirow}
\usepackage{soul}
\usepackage[utf8]{inputenc} 
\usepackage{float}
\usepackage{textcomp}

\DeclareMathOperator{\Tr}{Tr}

\newcommand{\mt}[1]{\mathcal{#1}}
\newcommand{\kv}{\textbf{k}}
\newcommand{\kb}{\bar{k}}
\newcommand{\qb}{\bar{q}}
\newcommand{\qv}{\textbf{q}}
\newcommand{\al}{\alpha}
\newcommand{\ep}{\epsilon}
\newcommand{\bt}{\beta}

\newcommand{\lmb}{\lambda}

\newcommand{\s}{\sigma}
\newcommand{\Sg}{\Sigma}
\newcommand{\ur}{\uparrow}
\newcommand{\dg}{\dagger}
\newcommand{\dr}{\downarrow}

\newcommand{\hb}{4Hb-TaS$_2$\xspace}

\newcommand{\AK}[1]{#1}
\newcommand{\AKn}[1]{#1}

\begin{document}
\title{Interlayer Charge-Transfer Ferroelectric Fluctuations as a Pairing Mechanism in van der Waals Superconductors }

\author{Ankan Biswas}
\thanks{These authors contributed equally to this work}
\affiliation{Physics Department, Ariel University, Ariel 40700, Israel}
\affiliation{Department of Physics, Bar Ilan University, Ramat Gan 5290002, Israel}

\author{Jagannath Sutradhar}
\thanks{These authors contributed equally to this work}
\affiliation{Physics Department, Ariel University, Ariel 40700, Israel}
\affiliation{Department of Physics, Bar Ilan University, Ramat Gan 5290002, Israel}
\affiliation{Department of Physics, Ghent University, Krijgslaan 281, 9000 Gent, Belgium}
\email{sutradj@biu.ac.il}
\author{Sudip Kumar Saha}
\affiliation{Physics Department, Ariel University, Ariel 40700, Israel}
\affiliation{Department of Physics, Bar Ilan University, Ramat Gan 5290002, Israel}
\author{Avraham Klein}
\affiliation{Physics Department, Ariel University, Ariel 40700, Israel}
\affiliation{Department of Physics, Bar Ilan University, Ramat Gan 5290002, Israel}
\email{avraham.klein@biu.ac.il}
\author{Jonathan Ruhman}%
\affiliation{Department of Physics, Bar Ilan University, Ramat Gan 5290002, Israel}
\email{jonathan.ruhman@biu.ac.il}

\date{\today}

\begin{abstract}
{
Signatures of unconventional superconductivity have been reported in a wide range of van der Waals (vdW) materials. 
However, their microscopic origin remains unclear due to competing electronic orders, strong spin--orbit coupling, and structural instabilities in the normal state. 
Here we investigate the role of interlayer breathing and shear modes in superconducting vdW heterostructures. 
Contrary to conventional wisdom---which assumes that weak interlayer bonding and large layer separation suppress electronic coupling to these modes---we show that the associated charge transfer can generate a substantial pairing interaction. 
We develop a theory of superconductivity mediated by such interlayer modes and demonstrate that proximity to a ferroelectric or antiferroelectric quantum critical point provides a strong-coupling pairing channel. 
Within a two-dimensional model with SU(2) symmetry and in-plane isotropy, we find an accidental degeneracy between interlayer triplet states, which can occur even for an $s$-wave in-plane gap. 
We further show that Josephson coupling between layers, arising from 
either static magnetism or induced by paramagnetic correlations,
can stabilize a time-reversal-symmetry-breaking superconducting state of the $s+i\,s$ type, which couples to magnetization when at least two mirror symmetries are absent.  
Our results are directly applicable to candidate chiral vdW superconductors such as \hb\ and to sliding ferroelectric metals, exemplified by bilayer MoTe$_2$. 
More broadly, our work identifies ferroelectric fluctuations as a promising route to unconventional pairing in vdW systems and motivates experimental searches for chiral multicomponent superconductivity.
}
\end{abstract}

\maketitle
\section{\label{sec:level1}Introduction}
The origin of unconventional phenomena in superconductors remains an important open question in condensed matter physics. These phenomena can arise from several mechanisms, including spin-orbit coupling, strong electronic correlations, and Coulomb interactions. Central to the problem is the nature of the pairing glue---the bosonic mediator of the attractive interaction responsible for Cooper pairing. In conventional superconductors, such as elemental metals and alloys, longitudinal acoustic phonons couple to the electronic density and typically produce a conventional superconducting state. In contrast, if the pairing boson transforms nontrivially under a symmetry, unconventional pairing states with richer structure can emerge.

Two-dimensional superconductors based on van der Waals (vdW) materials provide a versatile platform for investigating the emergence of unconventional superconductivity. Their reduced dimensionality, tunability, and material diversity enable access to a broad range of regimes, from SU(2)-symmetric to strongly spin-orbit-coupled systems~\cite{RevModPhys.96.021003, Saito_NatPhys2016, Cai_NatCom2021, Zhang_Nat2023, Bobkov_PRB2024, Holleis_2025}, and from weakly interacting, BCS-like superconductors to correlated insulators and non-s-wave states~\cite{Xi_NatPhy2016, PhysRevB.96.220506, PhysRevMaterials.2.094001, Menard_NatCom2017, Liao_NatPhys2018, PhysRevB.96.041201, You_PRB2021}. 
These systems exhibit rich phase diagrams that frequently feature competing orders other than superconductivity, including charge density waves~\cite{Fumega_NanLet2023, Coelho_CDW2019, Amir_npjQM2025}, correlated insulators, valley and magnetic order~\cite{Wang_ACSNano2022, Li_SciAdv2019}, and structural transitions~\cite{PhysRevB.105.104105, ma16010454, PhysRevB.107.134118, Liu_SciRep2020, PhysRevB.102.060103, Cheon_ACSNano2021}. Insights from other well-studied unconventional superconductors, such as the high-temperature cuprates, suggest that fluctuations near quantum phase transitions, controlling such rich phase diagrams, can play a central role in pairing and are likely the driving force in high-$T_c$ superconductivity. 

One interesting problem, both from the viewpoint of the fundamental physics and due to its many promising applications, is the role of ferroelectric (FE), i.e. inversion-breaking, fluctuations and distortions \cite{Cohen1967,Chaudhary2024,Kopasov2024,Palle2024}. In the context of vdW materials,
these are typically interlayer phonon modes breaking out-of-plane inversion. Naively, one would expect that the large interlayer separations in such compounds would imply only weak phonon-induced coupling, as is evidenced by the fairly small layer-induced band-splitting found in these compounds. However, such analysis neglects the possible role of \emph{interlayer charge transfer} due to the phonon modes, which renders them far more insidious. The coupling between interlayer phonon displacements
and FE polarization thus motivates a detailed study of their impact on superconductivity.

\begin{figure}
    \centering
    \includegraphics[width=0.9\hsize,clip, trim=50 25 0 25]{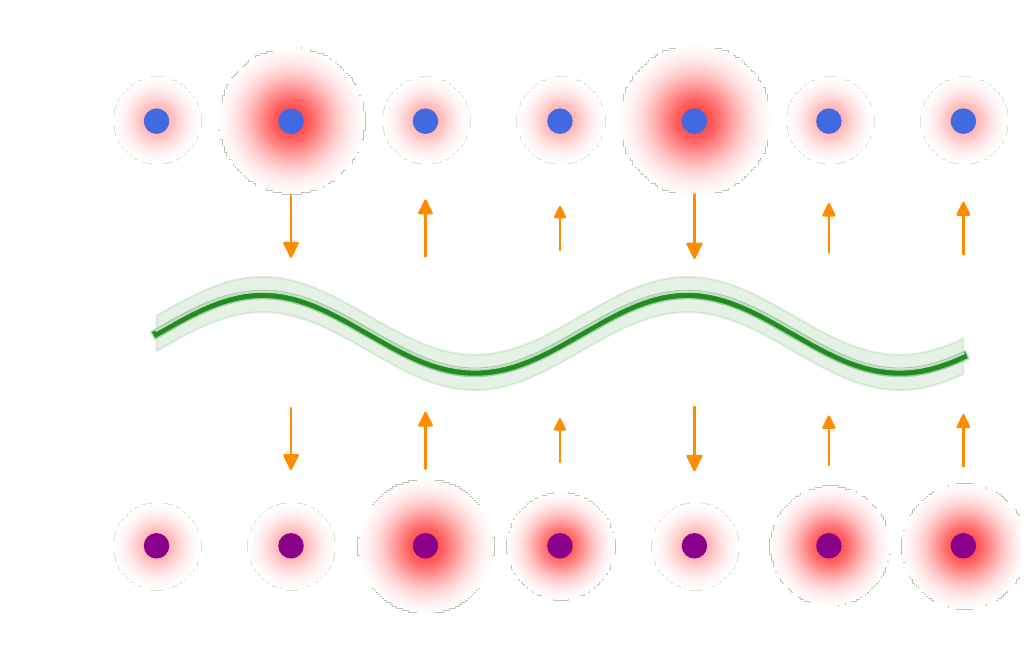}
    \caption{Schematic of the dynamical ferroelectric charge-transfer mechanism in a layered vdW system. A soft interlayer phonon  (green wavy pattern) produces instantaneous  displacements that locally break inversion symmetry  
    between the top and bottom electronic layers (blue and purple sites). Such inversion breaking can occur from a variety of motions, e.g. intermediate insulating layers such as occurs in \hb, or shear motion of the layers such as occurs in sliding ferroelectrics. The distortion mediates charge transfer (red clouds), and the
    resulting layer-odd charge imbalance generates local polarization dipoles (orange arrows), whose sign follows the  phonon distortion. 
    }
    \label{fig:illustration}
\end{figure}

Several notable mechanisms for such distortions exist.
%
%
One example is provided by sliding ferroelectrics (FEs)~\cite{FE_NatRev2023}--few-layer vdW crystals that can shift relative to one another, generating crystallographic misalignment and an accompanying out-of-plane FE polarization. Another 
is the 1T to 1T$^{\prime}$ structural transition observed in materials such as WTe$_2$~\cite{Duerloo_NatCom2014, PhysRevB.102.060103}, which can be tuned by pressure and is associated with a FE-like distortion. 
%
%
%
%
A
third, even more complex 
example, intimately related to and generalizing the previous ones, arises in metallic transition-metal dichalcogenides (TMDs) such as 4Hb-TaS$_2$, misfit compounds like (SnS)$_{1.15}$(TaS$_2$)~\cite{Misfit_ChemMat2022, WIEGERS1995152, Itahashi2025}, and 2H-TaS$_2$ intercalated with chiral molecules. 
In these systems, charge transfer due to differences in work function between layers can induce softening of the out-of-plane FE modes. An illustration of the basic idea is depicted in Fig. \ref{fig:illustration}.

Interestingly, superconductivity in these compounds is almost ubiquitous. 
Furthermore, understanding the pairing mechanisms in these superconductors is an especially interesting question given that they show signatures of non-s-wave pairing, including magnetic memory, edge states, anisotropy in $H_{c2}$, $\pi$-phase shifts in the Little-Parks effect~\cite{Amit_NatCom2021}, V-shape tunneling density of states~\cite{JeongNodalSC}, and anomalous thermal transport properties. 
This further underscores the importance of understanding the role of soft FE fluctuations in superconductivity within vdW materials.

In this paper we explore the role of the interlayer FE
mode in superconductivity in layered vdW materials. 
We focus especially on the region of ``soft'' 
FE 
fluctuations 
due to a nearby quantum critical point (QCP).
We construct a universal minimal model of two layers coupled by an interlayer phonon, and analyze its pairing tendencies. Our first main finding
%
is the existence of an accidental degeneracy between two leading \emph{layer-triplet} modes, that is only very weakly broken in the ordered state. In the presence of magnetic correlations, this degeneracy provides a natural avenue for spontaneous breaking of time reversal symmetry. The breaking is due to interlayer currents despite the fact that the in-plane gaps can be featureless s-wave. Our second main point is that the model can naturally be applied to 4Hb-TaS$_2$, as well as sliding FEs. In the case of 4Hb-TaS$_2$ the physical mode is actually an 
\emph{anti-FE} one, and we find that the nonlinear charge-transfer 4Hb-TaS$_2$ drives a significant softening of the interlayer mode, further motivating it as a possible SC mechanism.

In our work, for simplicity, we assume that the only fluctuating channel is this mode. In practice, of course, that is not the case, and presumably SC is driven by several phonon modes. However, as long is our mechanism is important, even in the presence of other phonon modes, it can be expected to be instrumental in selecting the preferred SC state. Accordingly, we do not explore too heavily the dynamical aspects of the QC fluctuations. For similar reasons, we also limit ourselves to s-wave in plane states, whereas in practice, certainly in the presence of complext polar-elastic or polar-magneto effects, the FE mechanism can also assist in creating non s-wave in-plane order.




\section{Model}
\label{sec:model}
We begin by formulating the model and subsequently discuss its physical relevance. For our model we use the simplest configuration with a nontrivial interlayer mode, namely two identical layers of itinerant fermions.
The layers are coupled via a minimal relevant phonon mode:
an 
optical phonon that breaks inversion along 
the out-of-plane ($\hat{z}$) direction, which mediates an interlayer charge transfer. 

The total action describing the system comprises three components and is written as:
\begin{equation}
S=S_{\psi}+S_{u}+S_{\psi u}.
\label{Eq:actions}
\end{equation}
The electronic part of the action is given by:
\begin{equation}
S_{\psi}=\Tr \sum_{\kb} \bar{\psi}_{\kb} \big(i k_{0}- 
\epsilon(\kv)
\big) \tau_0\s_0 \psi_{\kb}
,
\label{Eq:S_psi}
\end{equation}
where $\psi_{\kb}=\psi_{l,\al}(\kb)$ has four components, with two spin states $\al=\ur,\dr$ and two layer indices $l=t,b,$ representing the top and bottom layers. Here, $\s$ and $\tau$ are Pauli matrices in the spin and layer spaces, respectively. The four vector $\kb=(k_{0},\kv)$ combines the fermionic Matsubara frequency $k_{0}$ and the in-plane momentum $\kv=(k_x,k_y)$. We use the shorthand notation $\sum_{\kb}\equiv (\frac{a}{2\pi})^2 \sum_{k_0} \int d^2\kv$, with $a$ being the lattice constant. Finally, $\epsilon(\kv)$ is the dispersion, and for simplicity we assume a single parabolic band, $\epsilon (\kv) = \kv^2/{2m^*} - \mu$, where $m^*$ is the effective electron mass.

The phononic part of the action is given by :
\begin{equation}
S_u= {D_0^{-1}} \sum_{\qb} u_{\qb}\left(\frac{a^2}{c^2}q_{0}^2+ \textbf{q}^2 a^2+\frac{a^2}{c^2}m_0^2\right) u_{-\qb} 
,
\label{Eq:S_u}
\end{equation}
where $u_{\qb}$ is the phononic dimensionless displacement field along $z$ and  $\qb =(q_{0},\qv)$ is a four vector. $m_0$ is the bare mass, {$c$ is the phonon velocity, and $D_0^{-1}=\rho c^2 /2 $, where $\rho$ is a 2D  mass density associated with the phonon mode.}
$q_{0}$ is the bosonic Matsubara frequency. We again neglect the (typically quite significant) in-plane phonon anisotropy to keep things simple.

The coupling between the electrons and phonons is described by:
\begin{equation}
S_{\psi u}=\lmb \Tr \sum_{\kb,\qb} u_{\qb} \bar{\psi}_{\kb} \tau_z\s_0 \psi_{\kb+\qb}\, ,
\label{Eq:S_psi_u}
\end{equation}
where $\lmb$ is the coupling strength. Explicitly expanding the layer indices, we see that the  electronic bilinear coupling to the phonons is $(\bar{\psi}_{t} \psi_{t}-\bar{\psi}_{b} \psi_{b})$ (suppressing spin and momentum indices for clarity).  This corresponds to the difference in charge density between the two layers and $\bar\psi\tau_{z}\psi$ transforms like a dipole moment in the $\hat{z}$ direction. We note, that another  inversion breaking coupling is allowed, namely a Rashba spin-orbit like one, $\sim u \bar{\psi}\tau_0\kv\times\boldsymbol{\sigma}\psi$, which has been shown to be relevant in a variety of materials \cite{RSOC_BordoloiJAP2024, Saha2025Strong,venditti2025spin,nawwar2025large,gastiasoro2022theory,Fu2015,KleinKoziiRuhmanFernandes2023QuantumFerroelectricMetals,kozii2015odd}. However, in the layered vdW compounds we are considering, strong Ising SOC locks the spins in the out-of-plane direction, and this reduces the expected contribution from the Rashba term. In any case, in the current work we neglect it for simplicity, but we will discuss its possible impact in the final section.

\section{The pairing problem}

\subsection{Pairing symmetry}
\label{Sec:pair_sym}
The form of the electron-phonon coupling $S_{\psi u}$ determines the structure of the superconducting pairing function.
Specifically, the vertex factor $\tau_z$ from $S_{\psi u}$ enters the pairing diagram shown in Fig.~\ref{Fig:F_diagram}(c). {We can draw a parallel between this coupling and that of a magnetic interaction, which is known to prefer spin triplet pairing~\cite{PhysRevLett.30.1108,Chubukov2003first}. In this case the layer index takes the role of spin leading to a \emph{layer triplet} state.
This specific type of coupling, $\tau_z$, gives rise to two attractive and one repulsive triplet channel~\cite{Fay1980, Santi2001, Chubukov2004}. }

To see this, assume for simplicity rotational invariance in both spin and real space, implying that the spin and layer index are decoupled. Then we can decompose the pairing functions into channels
\begin{equation}
    \Phi(\kb) = i\sigma_y\sum_{j=0,x,z}\tau_j \Phi_{j}^{sing} + i\tau_y\sum_{j=x,y,z} i\sigma_y\sigma_j\Phi^{trip}_j\,,
\end{equation}
where we assume an s-wave pairing gap in the plane $\Phi(\mathbf{k})=const$. 
We can infer the possible pairing channels directly from the linearized  gap equation [Fig.~\ref{Fig:F_diagram}(c) and Eq.~\eqref{Eq:gap_eq}]. To have a non-trivial solution we require both sides of the equation to have the same sign. Because each interaction vertex in Fig.~\ref{Fig:F_diagram}(c) carries a factor of $\tau_z$, the kernel acts on a basis element $\tau_j\Phi_j$ as $ \tau_z\tau_j\tau_z= \pm\tau_j$. 
This condition is met only for the layer-triplet components $\tau_z$ and $\tau_0$.
In contrast, the layer-singlet $i\tau_y$ and the layer-triplet $\tau_x$ channels acquire a minus sign,
and therefore correspond to repulsive channels in the linearized gap equation.


At the superconducting transition temperature $T_c$, the two channels $\tau_z$ and $\tau_0$ are degenerate. Consequently, superconducting pairs are formed within individual layers, with no interlayer pairing component.
To see this, note that the two degenerate channels 
can be linearly combined to construct layer-specific superconducting order parameters: $\Phi_t=(\Phi_z+\Phi_0)$ and $\Phi_b=(\Phi_0-\Phi_z)$,  corresponding to the top and bottom layers. The degeneracy of these states opens a path to break  time-reversal symmetry ($\mt{T}$) in the form of an ``$s+i\,s$'' order parameter  
\begin{equation}
\Phi_t+e^{i\phi}\Phi_b = e^{i\phi/2}\left(\cos{\phi\over2} \,\Phi_0 - i \sin {\phi\over 2}\, \Phi_z\right)\,,
\end{equation}

where $\phi$ is the relative phase between the two layers. A possible scenario that energetically prefers such a state is discussed in Sec.~\ref{SSec:TRSbrokenSC}.

\subsection{Solutions of the gap equation}
\label{Sec:sol_gap} 
To make our theory as general as possible we will formulate our computations to account for the quantum critical fluctuations and the possibility of strong coupling superconductivity.
We begin by evaluating the self-energies of the electronic and bosonic fields in the normal state. These results are then used to solve the linearized gap equation to determine the critical temperature $T_c$. 

\begin{figure}
    \centering
    \includegraphics[width=1\linewidth]{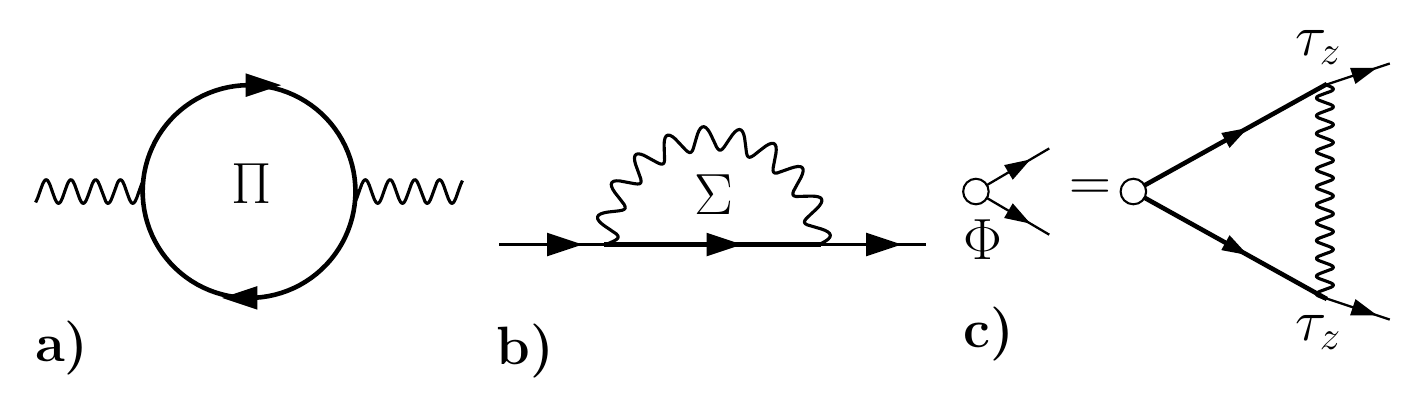}
    \caption{Feynman diagrams for (a) the bosonic self-energy, (b) the electronic self-energy, and (c) the linearized pairing equation.}
    \label{Fig:F_diagram}
\end{figure}

As shown in Fig.~\ref{Fig:F_diagram}(a), the bosonic self-energy is given by
\begin{equation}
\Pi(\qb)
=
-\bar{\lmb} T\Tr \sum_{\kb} \tau^z \s_0 G(\kb) \tau^z  \s_0 G(\kb+\qb)\, , 
\label{Eq:Pi}
\end{equation}
where $G(\kb)$ is the electronic propagator. We define the effective coupling $\bar{\lmb}\equiv \lmb^2 D_0$, where $D_0$, 
having dimensions of inverse energy density, is factored out of the bosonic propagator so that the latter is dimensionless.
We assume that $G$ has no nontrivial structure in the layer (i.e., $\tau$) space. 
In the limit $T\rightarrow 0$, the bosonic self-energy can be approximated as

\begin{equation}
 \begin{aligned}
\Pi(\qb) &\approx 4\bar{\lambda}\nu_F \left( 1-   \frac{\vert q_0 \vert}{v_F \vert \bm{q} \vert} \right),
 \label{eq:bubble_2D_LD1}
  \end{aligned}
 \end{equation}
assuming $q_0\ll v_F |\qv|$, and using the free fermionic propagator for $G$. The constant term renormalizes the bosonic mass to $m_r^2 = m_0^2 - 4 \bar{\lambda} \nu_F c^2/a^2$; where $\nu_F=m^{*} a^2/(2\pi)$
is the 2D density of states per spin at the Fermi level, $k_F$ and $v_F$ are the Fermi momentum and Fermi velocity respectively. The frequency and momentum-dependent part in Eq.~\eqref{eq:bubble_2D_LD1} leads to Landau damping of the bosonic mode.  
To characterize the distance from QCP, we introduce the dimensionless control parameter
\begin{equation}\label{eq:rdef}
  r = m_r^{2} \left(a/{c}\right)^2  .
\end{equation}

We first solve the pairing equation in the disordered phase $(r>0)$, and discuss the ordered phase $(r<0)$
in the next section.

The electronic self-energy [Fig.~\ref{Fig:F_diagram}(b)] is given by
\begin{equation}
\Sigma(\kb)
=
\bar\lmb T \sum_{\qb} \tau^z\s_0 G(\kb+\qb) \tau^z\s_0 D(\qb)\,,
\label{Eq:Sigma}
\end{equation}
where $D(\qb)$ is the dimensionless bosonic propagator. Away from the quantum critical point-- where the renormalized bosonic mass dominates the denominator of $D(\qb)$-- the electronic self-energy exhibits a fermi-liquid-like behavior: $\Sg\sim k_0$. However, when the bosonic mass vanishes near criticality, the self-energy acquires a non-Fermi liquid form with a characteristic power-law frequency dependence: $|\Sg|\sim |k_0|^{2/3}$. In practice, the non-Fermi liquid behavior is suppressed by the onset of superconductivity, which gaps out low-energy excitations.

\begin{figure*}
    \centering
    \includegraphics[width=1\linewidth]{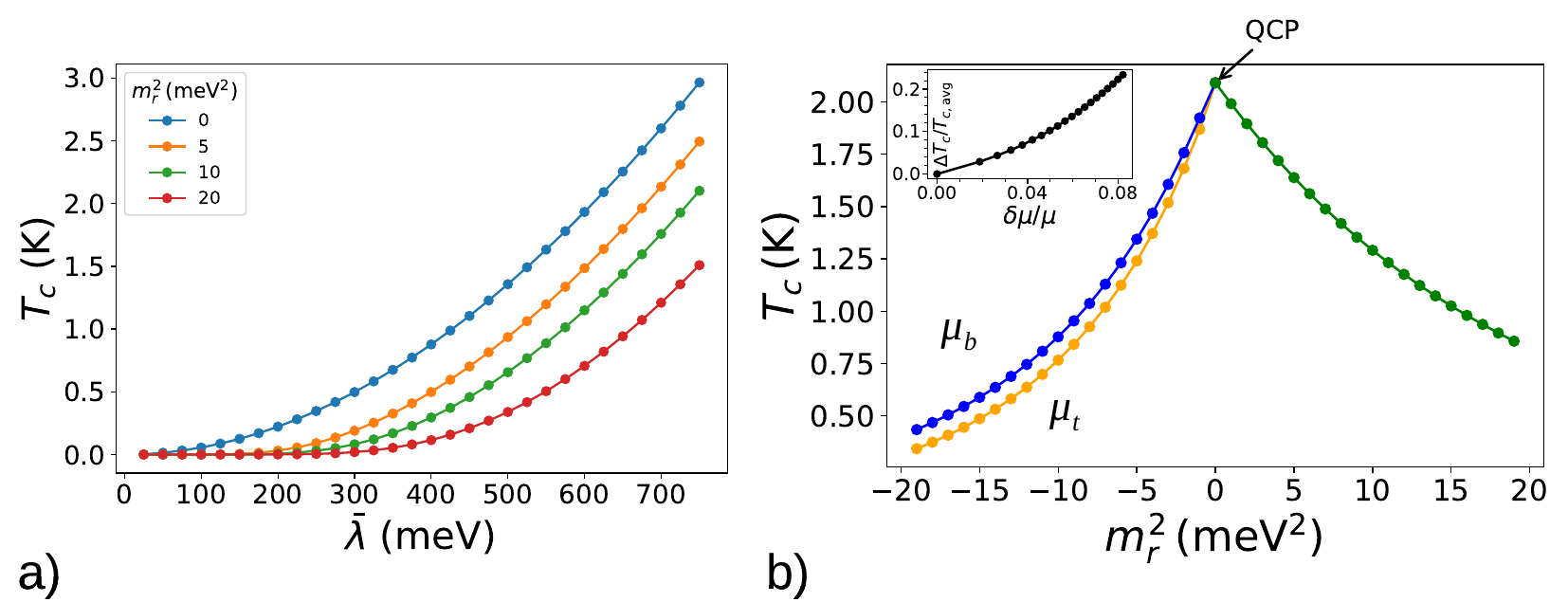}
    \caption{Superconductivity from interlayer charge-transfer soft fluctuations. The figures depicts $T_c$ obtained from the linearized gap equation [Eq.~\eqref{eq:eliashberg_gapeq_matrix},\eqref{eq:eliashberg_gapeq_matrix1} ] 
as a function of (\textbf{a}) coupling strength $\bar{\lambda}$ for several values 
of the renormalized mass $r>0$, and (\textbf{b}) $r$ at fixed coupling. 
\textbf{a)} Near the zero-mass limit, $T_c$ rises rapidly with increasing coupling. 
\textbf{b)} 
$T_c$ is maximal near the QCP 
and decreases on both sides away from criticality. 
For $r>0$ (disordered phase), $T_c$ follows the single green curve. 
For $r<0$ (ordered phase), the superconducting transition splits into two nearly 
degenerate branches: the blue curve for the bottom layer and the orange curve for the 
top layer.  This splitting is $\delta\mu=\lambda u_0$, with $\mu_t > \mu_b$ (see details in Appendix \ref{app:ordered}), is small near the QCP, 
and grows deeper into the ordered state. We used $\bar{\lambda} = 625$ meV for this Figure
\textbf{Inset:} Relative splitting of the transition temperatures, 
$\Delta T_c/T_{c,\mathrm{avg}}$, plotted as a function of the dimensionless chemical 
potential difference $\delta\mu/\mu$, showing a monotonic enhancement with 
increasing $\delta\mu$. We linearized the electronic dispersion near $\mu=E_F$, and used $u_0= 0.353|r|$, $E_F =\hbar^2 k_F^2/2m^*= 660$ meV, and $m^*=2m_e$ for our numerical result. 
}
    \label{Fig:Tc_coupling}
\end{figure*}

Using the Feynman diagram shown in Fig.~\ref{Fig:F_diagram}(c), the linearized pairing equation can be written as
\begin{equation}
\Phi(\kb) =\bar\lmb T \sum_{\qb} \tau^z\s_0 G(\kb+\qb)D(\qb)G(-\kb-\qb)  \Phi(\kb+\qb) \tau^z\s_0\,.
\label{Eq:gap_eq}
\end{equation}
Eq. \eqref{Eq:gap_eq}, supplemented with the electronic and bosonic self-energies obtained from Eqs.~\eqref{Eq:Pi} and \eqref{Eq:Sigma}, can be solved for the critical temperature $T_c$  
(see Appendix \ref{appA} for details).
We obtain $T_c$ as a function of the electron-phonon coupling strength. For concreteness, we use parameters relevant to the unconventional superconductor \hb, which will be discussed in detail below. The dependence on $\bar{\lambda}$ is shown in Fig.~\ref{Fig:Tc_coupling} (a).  $T_c$ increases as the phonon mass vanishes, as shown in Fig.~\ref{Fig:Tc_coupling}(b).   
For example, to achieve a critical temperature of
$T_c\sim 2$K, using a renomalized phonon mass $m_r\approx 2.2$ meV and a typical Fermi momentum $k_F \approx 0.6~\mathrm{\AA^{-1}}$ \cite{Kanigel_npj2024} , the required electron-phonon coupling strength ($\bar\lmb$) is $\sim 600$ meV [Fig.~\ref{Fig:Tc_coupling}(a)], where the effective electronic mass is taken to be $m^*=2 m_e$ (twice the bare electron mass). 
However, as the phonon mode softens, even a small electron-phonon coupling leads to a rapid enhancement of $T_c$.

In this paper we consider the interlayer FE phonon $u$ as the sole pairing mechanism. In practice this is not the case, as there are obviously other sources for pairing. For example, many vdW materials become superconducting without charge transfer fluctuations. 
For our results to be relevant it is sufficient that the contribution from these fluctuations will be large enough to select $\tau_0,\tau_z$ over competing channels. In practice, this means that incorporating $u$ into the pairing kernel shifts $\Delta T_c/T_c$ by order of unity. As we discuss below, this is the case for the material platforms we considered.


\subsection{Ordered FE state and layer splitting}
We now extend the theory discussed in Sec.~\ref{Sec:sol_gap} to the FE ordered state, i.e., $r <0$. 
In this phase, the FE displacement develops a static expectation value $\langle u\rangle= u_0 \neq 0$, 
obtained by minimizing the bosonic Ginzburg--Landau free energy. By symmetry, via Eq.~\eqref{Eq:S_psi_u}, a nonzero $u_0$ also gives rise to an out-of-plane polarization $P_z = -ed\langle\bar{\psi}\tau_z\psi\rangle\propto u$, where $e$ is the elementary charge and $d$ is the interlayer separation.
In this state, it is convenient to expand the phonon field into a static and a fluctuating part, 
$u_{\bar q} = u_0\, \delta_{\qv,0} + \delta u_{\bar q}$, where $\delta u_{\bar q}$ represents fluctuations around the ordered minimum.
The static component $u_0$ acts as a layer-odd potential that shifts the chemical potentials of the two layers  
by 
$\delta\mu = \lambda u_0$.
As a result
propagators become layer-resolved 
$G_{l}$, where $l=t,b$ denotes the top and bottom layers, respectively. For the rest of the paper, we will use $l$ as a suffix to indicate the layer specific quantities. The detailed form of the ordered-state action is given in the Appendix \ref{app:ordered}.

At the same time, the curvature of the free energy in the bosonic sector at the ordered minimum modifies the 
form of the phonon propagator $D(\bar q)$, effectively replacing $r$ by $2|r|$.
(see Appendix \ref{app:ordered} for details).
Compared with the disordered phase, the layer asymmetry introduced by $u_0$ produces small layer-dependent variations in the Fermi momentum ($k_{Fl}$), velocities ($v_{Fl}$), and for a non-parabolic band the DOS ($\nu_{Fl}$) as well. Crucially, the Ising $\tau^z$ form of the interaction in Eq. \eqref{Eq:S_psi_u} insures that there is no direct interlayer scattering even in the ordered state. As a result, the Eliashberg gap equations decompose into two independent, layer-resolved forms for the top and bottom layers. The overall kernel retains the same structure as in the disordered state, with only these layer-specific parameters distinguishing the two.

Substituting the ordered-state propagators into the one-loop bubble, Eq. \eqref{Eq:Pi} becomes (see more details in Appendix)
\begin{equation}
\begin{aligned}
\Pi(\qb) &= -2\bar{\lambda} T 
 \sum_{\kb; \,l = t,b}  G_l(\kb) G_l(\kb + \qb)\\
&\approx 4 \bar{\lambda} \nu_F \left(1- \frac{1}{v_{eff}}\frac{|q_0|}{|\mathbf{q}|}\right)\,. 
\end{aligned}
\label{eq:pi_order}
\end{equation}
where $1/v_{eff}= 1/2\sum_{l} 1/v_{F l}$. DOS ($\nu_{F}$) remains unchanged due to our parabolic model, Eq. \eqref{Eq:S_psi}
and $v_{Fl}$ are the split Fermi velocities. 
This simple form is a direct result of the diagonal $\tau_z$ coupling, which also renders the ordered-state fermionic self-energy diagonal
in the layer basis
\begin{equation}
\begin{aligned}
\Sigma_{l}(k_0)&
= \frac{-\,i\,\bar\lambda\,\nu_{F}}{k_{Fl}a}\;
T\sum_{q_0}\mathrm{sgn}\!\bigl(k_0+q_0\bigr)\,d(q_0)\, ,\label{eq:pairing-matsubara}
\end{aligned}
\end{equation}
where by $d(q_0)$ we denote the frequency dependent kernel which arises from the momentum-integrated bosonic propoagator in the limit of vanishing momentum transfer,
\begin{equation}
 \begin{aligned}
 d(q_0)&=\int_0^{\Lambda} D(\bar{q})\, d|\textbf{q}|\,. 
          \label{eq:boson_function_full3}
 \end{aligned}
\end{equation}
Near the QCP, this reproduces the non-Fermi-liquid scaling with layer-dependent prefactors. 
The ordered-state pairing equations are now,
\begin{equation}
\Phi_{l}(k_0)
=\frac{\bar\lambda\, \nu_{F}}{k_{Fl}a}\;
T\sum_{q_0}
\frac{\Phi_{l}(k_0+q_0)\,d(q_0)}{\bigl|\tilde\Sigma_{l}(k_0+q_0)\bigr|}\, .
\label{eq:paringdordered}
\end{equation}
Equation~\eqref{eq:paringdordered}, treated as an eigenvalue problem, can then be solved numerically
for $T_c$ across the entire phase diagram as a function of the distance  $r$ from the QCP,
once the following technical point is
addressed: the $q_0=0$ divergence in Eq.~\eqref{eq:boson_function_full3} must be handled. For the $\tau_z$ coupling, this can be accomplished in both ordered and disordered states
~\cite{MillisSachdevVarma1988Inelastic, AbanovChubukovNorman2008GapAnisotropy, MoonChubukov2010QuantumCritical, ChubukovAbanovWangWu2020InterplaySuperconductivity} 
by recasting the problem in terms
of the gap function, 
\begin{equation}
\Delta_{l}(k_0)=k_0\Phi_{l}(k_0)/\tilde\Sigma_{l}(k_0)\, ,
\end{equation}
(see Appendix for details and the resulting equations). As shown in Fig.~\ref{Fig:Tc_coupling}(b), the ordered phase ($r<0$) exhibits a split in the layer dependent $T_c$, while the disordered phase ($r>0$) yields the same $T_c$ for both layers.  However, this splitting is remarkably small, with 
\begin{equation}
    \frac{\Delta T_c}{T_c}  \propto r^2.
    \label{eq:deltaTc/Tc}
\end{equation}
(see details in Appendix \ref{app:delta}). 
The small splitting is a generic feature of SC arising from a soft mode, which involves small momentum transfer fluctuations~\cite{Klein2018, Klein2019,KleinKoziiRuhmanFernandes2023QuantumFerroelectricMetals}.

\subsection{$\mt{T}$ broken SC state}
\label{SSec:TRSbrokenSC}
One of the most intriguing predicted features of unconventional superconductors is the spontaneous breaking of TRS, which may occur at or below $T_c$, either as an intrinsic instability of the superconducting state or in the presence of preexisting magnetic order. A hallmark of TRS-breaking superconductivity is the appearance of spontaneous internal magnetic fields, which can be probed by muon spin relaxation~\cite{LukeTRSSr2RuO4_nat1998, RibakChiralSC_SciAdv2020} and polar Kerr effect measurements~\cite{XiaPolarKerrSr2RuO4_PRL2006, Kapitulnik_2009, LevensonPolarKerr_PRL2018}.

In the absence of strong ferromagnetism in the normal state, TRS-breaking in superconductors typically requires the existence of multiple nearly degenerate pairing channels, arising  from a multi-component order parameter. As discussed in Sec.~\ref{Sec:pair_sym}, the near degeneracy between the $\tau^0$ and $\tau^z$ pairing channels in our model naturally allows for a TRS-breaking superconducting state. The small energetic splitting found in the preceding section suggests that fluctuations driving such a state may remain relevant even in the ordered phase.

We now analyze the consequences of TRS breaking for the superconducting order parameter in this system. The breaking of time-reversal symmetry permits a finite relative phase $\phi$ between the intralayer superconducting order parameters $\Delta_t$ and $\Delta_b$, corresponding to a complex superconducting state characterized by $\arg(\Delta_t \Delta_b^*)=\phi$. While TRS-breaking $s+i\,s$ states have been extensively discussed in multiband superconductors~\cite{StanevTes2010,FernandesMaitiWolfleChubukov2013,GrinenkoPRB2017,GrinenkoArxiv2018,GrinenkoNaturePhys2020}, in those cases the distinct $s$-wave components typically reside in different momentum-space bands and preserve all spatial symmetries, resulting in the absence of symmetry-enforced bulk supercurrents.

In contrast, in the present system the two $s$-wave order parameters $\Delta_t$ and $\Delta_b$ are associated with spatially separated two-dimensional layers. As a result, a nontrivial relative phase $\phi$ generates an interlayer Josephson current whose direction is determined by the sign of $\phi$. These currents form circulating real-space current loops, leading to spontaneous orbital magnetization and internal magnetic fields. Consequently, the TRS-breaking superconducting state in this geometry is intrinsically magnetically active and breaks both time-reversal and inversion symmetries.

To show this we characterize the TRS-breaking order using the pseudoscalar quantity
\begin{equation}
\mathcal K
= i\,u_0\left(\Delta_t\Delta_b^*-\Delta_t^*\Delta_b\right)\, ,
\end{equation}
where $u_0$ denotes the ferroelectric order parameter. The quantity $\mathcal K$ is odd under TRS and even under all remaining spatial symmetries, and therefore transforms as a TRS-odd pseudoscalar. In this sense, $\mathcal K$ shares the same transformation properties as a magnetic charge density, although it does not correspond to a true magnetic monopole.

When \emph{at least two mirror symmetries are broken}, symmetry permits a linear coupling between $\mathcal K$ and a pseudovector such as the magnetization $\boldsymbol{\mt M}$,
\begin{equation}
E_{\mathcal T}
= -c_1 \,(\boldsymbol{\mt M}\cdot\hat{\boldsymbol n})\,\mathcal K\, ,
\label{eq:trs}
\end{equation}
where $\hat{\boldsymbol n}=\hat{\boldsymbol m}_1\times\hat{\boldsymbol m}_2$ is a symmetry-allowed axial direction determined by the broken mirror planes and $c_1$ is a constant. This coupling implies that a finite magnetization induces a relative phase between the two layers, and conversely that a TRS-breaking interlayer phase difference generates a spontaneous magnetization along $\hat{\boldsymbol n}$. Even in the absence of a net magnetization, strong magnetic fluctuations can lower the free energy through this coupling, favoring the emergence of a TRS-breaking superconducting state. Note, that if $\mathcal{M}$ is in the plane, then $u_0\hat{z}$ itself serves as one of the broken mirror planes.

In order to provide a concrete mechanism in addition to our qualitative arguments, we 
analyze the interlayer Josephson coupling generated by single-electron tunneling and show under which circumstances it can favor TRSB.
We consider interlayer hoppings of the form 
\begin{equation}
H_{tb} = \sum_{i=x,y}\sum_{j=0,x,y,z}\Lambda_{ij} \sum_{\kv} {\psi}^{\dg}_{\kv,l\al} \tau^i_{ll'}\s_j^{\al\bt} \psi_{\kv,l'\bt}\, .
\end{equation}
where $\al, \bt$ denote spin indices, and $\Lambda_{ij}$ encode the symmetry-allowed hopping channels. Their transformation under inversion ($\mt{I}$)
and time reversal $(\mt{T})$ is summarized in Table~\ref{Tab:tausigma}.
Throughout, we assume the interlayer hopping is sufficiently weak that it does not alter the intrinsic intralayer pairing channels ($\tau_z$ and $\tau_0$); the opposite limit is discussed later in Sec.~\ref{Sec:Interlayerhopping}.

\begin{table}[t]
\centering
\begin{tabular}{|c|c|c|}
\hline
Hopping matrices  & \makecell{Inversion \\ $\mt{I}$} & \makecell{Time Reversal \\ $\mt{T}$} \\
\hline \hline
$\tau_x\sigma_x$, $\tau_x\sigma_y$, $\tau_x\sigma_z$  & + & - \\
\hline
$\tau_x\sigma_0$  & + & + \\
\hline
$\tau_y\sigma_x$, $\tau_y\sigma_y$, $\tau_y\sigma_z$  & - & + \\
\hline
$\tau_y\sigma_0$  & - & - \\
\hline
\end{tabular}
\caption{“$+$” (“$-$”) indicates that the corresponding interlayer hopping matrix is even (odd) under inversion or time reversal. }
\label{Tab:tausigma}
\end{table} 

Within a second order perturbation analysis we act with 
$\psi^\dg_{l\bt}\psi_{\bar{l}\al}$ or $\psi^{\dg}_{\bar{l}\bar\al} \psi_{l\bar\bt}$, which are time-reversed hermitian conjugates, on a (spin-signlet) superconducting ground state creates the \textit{same} two-quasiparticle intermediate state--one quasiparticle in each layer with spins 
$\bar\al$ and $\bt$ (see Appendix \ref{sec:app-jos}). Here, $\bar{l}$ is the opposite layer of $l$ and $\bar\al(\bar\bt)$ the opposite spin of $\al(\bt)$. The interference between the two hopping paths yields the following correction to the energy of the system
\begin{equation}
\delta E_2 \propto -\mathrm{sgn}(\al\bt) [M_{l\al}^{\bar{l}\bt} (M_{\bar{l}\bar\bt}^{l\bar\al})^* \Delta_t \Delta_b^* + c.c]\, , 
\label{Eq:josephson_energy}
\end{equation}
where $M_{l\al}^{\bar l\bt}=\sum_{ij}\Lambda_{ij}\tau_i^{l\bar l} \s_j^{\al\bt}$ is the hopping matrix element for $l\al\to \bar{l}\bt$. This is precisely a term of the form of Eq. \eqref{eq:trs}, with $\mt{M} \mt K\propto- iM_{l\al}^{\bar{l}\bt} (M_{\bar{l}\bar\bt}^{l\bar\al})^*$. 

When both $\mt{T}$ and $\mt{I}$ are broken and spin-flip channel dominates, one finds, for example, 
\[
\begin{aligned}
M_{t\ur}^{b\dr}=\Lambda_{xx}-\Lambda_{yy}+i(\Lambda_{xy}+\Lambda_{yx})
\,,\\
M_{b\ur}^{t\dr}=\Lambda_{xx}+\Lambda_{yy}+i(\Lambda_{xy}-\Lambda_{yx})\,.
\end{aligned}
\]
In general, define the relative phase between the two interfering hopping amplitudes and between the order parameters by
\[
M_{t\ur}^{b\dr} (M_{b\ur}^{t\dr})^*= e^{-i\phi_h}|M|^2\,, \quad
\Delta_t \Delta_b^*=e^{i\phi}|\Delta|^2\,.
\]
Then
\begin{align}
    \delta E_2 \propto \mathrm{cos}(\phi-\phi_h) |M\Delta|^2\,,
\end{align}
which is minimized for $\phi=\phi_h+\pi$. Thus, the microscopic phase $\phi_h$
set by the interlayer hopping channel selects the preferred interlayer relative phase $\phi$. In the symmetric limit,  $\Lambda_{xx}=\Lambda_{yy}=\Lambda_{xy}=\Lambda_{yx}$, one has $\phi_h=\pi/2$, so the minimum occurs at $\phi=-\pi/2$ (a complex relative phase). When one of the symmetries, $\mt{T}$ or $\mt{I}$ is broken, a nematic phase ($s\pm s$) is preferred. 
If spin-preserving hopping dominates, the sign structure differs, giving
$\phi=\phi_h$. If spin-flip and spin-preserving channels contribute equally, the quadratic terms leave 
$\phi$ undetermined, as $\delta E_2=0$; higher-order processes are then required to fix $\phi$.
The symmetry determined values of
$\phi_h$ are summarized in Table~\ref{Tab:phi_h}. 
The case with mirror-$z$ symmetry is similar to the findings above and 
is discussed in the Appendix. 

We conclude that an interlayer Josephson mechanism, combined with broken $\mt{T}$ and $\mt{I}$, naturally explains the emergence of complex SC pairing of the form $s + e^{i\phi}s$.

\begin{table}[t]
\centering
\begin{tabular}{|c|c|cc|}
\hline
\makecell{Inversion \\ $\mt{I}$} & \makecell{Time Reversal \\ $\mt{T}$} &  \multicolumn{2}{c|}{Relative Phase $\phi_h$} \\
\cline{3-4}
    &       & \multicolumn{1}{c|}{ \makecell{Spin Flip \\ $\phi_h=\phi-\pi$} } & \makecell{No Flip \\$\phi_h=\phi$} \\
\hline\hline
+ & - & \multicolumn{1}{c|}{0}        & \makecell{$0$ if $\Lambda_{xz}<\Lambda_{x0}$ \\ $\pi$ if $\Lambda_{xz}>\Lambda_{x0}$}    \\
\hline
- & + & \multicolumn{1}{c|}{$\pi$}    & 0         \\
\hline
- & - & \multicolumn{1}{c|}{$\phi_h$} & $\phi_h$  \\
\hline
+ & + & \multicolumn{1}{c|}{$--$}       & 0         \\
\hline
\end{tabular}
\caption{Symmetry constraints on the microscopic phase $\phi_h$ entering Eq.~\eqref{Eq:josephson_energy}, for spin flip and no flip hopping processes. ``$+$" (``$-$") indicates the symmetry is preserved (broken). ``$--$'' denotes that the corresponding hopping process is forbidden by symmetry. An entry of $\phi_h$ denotes a symmetry-unconstrained phase.}
\label{Tab:phi_h}
\end{table}
\subsection{Interlayer hopping and pairing symmetry}
\label{Sec:Interlayerhopping}

We conclude by briefly discussing when 
interlayer hopping is strong enough to modify the SC pairing symmetry. The electron phonon coupling in Eq.~\eqref{Eq:S_psi_u} is generalized to:
\begin{equation}
S^\prime_{\psi u} = \sum_{i=x,y,z}\lmb_i \sum_{\kb,\qb} u_{\qb} \bar{\psi}_{\kb} \tau_i\s_0 \psi_{\kb+\qb}\,.
\label{Eq:S_psi_u2}
\end{equation}
By symmetry, the term with $\lambda_x$ is odd under inversion $\mt{I}$, while that with $\lambda_y$ is odd under both $\mt{I}$ and $\mt{T}$. The two degenerate superconducting channels in this case are $\Phi_0\propto\tau_0$ and $\Phi_\lmb\propto\hat\lambda\cdot\vec\tau$, where the vector $\bm{\lmb}=(\lmb_x,\lmb_y,\lmb_z)$ picks out a direction $\hat{\lmb}=\bm\lmb/|\bm\lmb|$ in layer space. 
In general, the pairing cannot be written as independent layer-resolved orders: interlayer components mix with intralayer ones ($\tau_x$ and $\tau_y$). Moreover, the interlayer–intralayer mixing enables an unconventional pairing state with mixed spin-singlet and spin-triplet character. 

The interlayer pairing competes strongly with the FE order~\cite{Jindal2023}. Inside the FE phase, an interlayer chemical potential difference splits the two bands at the Fermi energy—effectively acting as a Zeeman field in the layer space. This naturally hinders interlayer pairing and biasses the SC order toward a layer-triplet structure  $\Phi_0$. In contrast, near the FE quantum critical point, the splitting is small, making the two channels $\Phi_0$ and $\Phi_{\lmb}$ nearly degenerate. Thus one may expect a change of pairing symmetry when tuning from the vicinity of the QCP into the deep FE phase.  

The Josephson coupling argument that locks the phase between degenerate SC orders breaks down for interlayer pairing with mixed symmetry.

\section{Application to specific materials}

So far, the discussion based on our toy model has been quite general and applicable to a broad class of vdW materials. We now turn to material-specific scenarios that motivate the phonon coupling with electric dipole moment [Eq.~\eqref{Eq:S_psi_u}].
\subsection{Interlayer breathing modes -- application to 4H\lowercase{b}-T\lowercase{a}S$_2$}
\begin{figure}[t]
    \centering
\includegraphics[width=\linewidth]{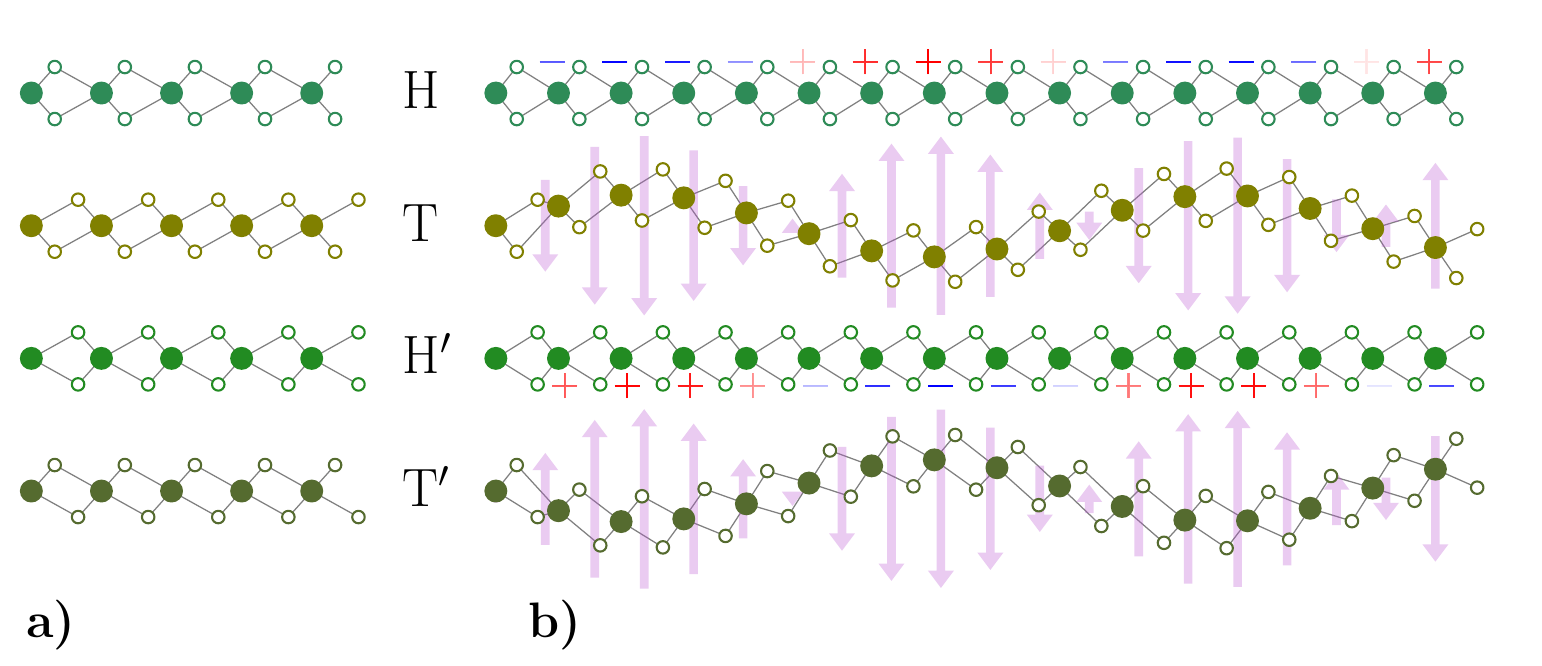}
    \caption{(a) A cross-sectional view along the \textbf{a}-axis of a 4Hb-TaS$_2$ unit cell \cite{RibakChiralSC_SciAdv2020} shows alternating H and T layers, with solid ({\scriptsize\CIRCLE}) and open ({\scriptsize\Circle}) circles representing Ta and S atoms, respectively. (b) Excitation of an $\hat{z}$-polarized optical phonon ($\cdot$ $\rightarrow$ $\cdot$ $\leftarrow$, see Table \ref{Tab:phonon_modes}) produces out-of-phase displacements in the two T-layers. This effectively transfers charge between the H-layers. Because the transferred charge depends strongly on the H–T interlayer distance, the oscillating T-layer motion causes the direction of charge transfer to alternate (indicated by $+$ and $-$). This process generates an oscillating dipole moment, producing antiferroelectric fluctuations represented by the arrow, whose length is proportional to the polarization. In the minimal model described by Eq.~\eqref{Eq:actions}, the two electronic bands represent the two H-layers, while $u_{\bar{q}}$ corresponds to the T-layer displacement.
     }
    \label{Fig:4Hb_structure}
\end{figure}

In recent years, \hb has attracted considerable attention due to its rich normal state behavior and unconventional superconductivity~\cite{RibakChiralSC_SciAdv2020,Amit_NatCom2021,Silber2024,nayak2021evidence}. This material consists of alternating layers of 1T-TaS$_2$ and 1H-TaS$_2$. The stacking sequence along the $\hat{z}$-direction follows a four-layer periodicity: H-T-H$^\prime$-T$^\prime$, where the unprimed and primed layers serve as inversion partners, as shown in Fig.~\ref{Fig:4Hb_structure}(a). In addition, both H and T undergo charge-density-wave deformations, with the 1T layer forming ``star-of-David'' 13 atom supercells. The 1T layers are believed to host strong magnetic correlations,
while the 1H layers remain metallic.

One intriguing aspect of the physics in this material is interlayer charge transfer.
Both density functional theory calculations~\cite{kumar2023first,Crippa_ncom2024} and experimental observations~\cite{kumar2023first,Kanigel_npj2024} indicate a charge transfer from the 1T layers to 1H, which is strongly dependent on the interlayer separation.
This feature is captured by our model through the effective electron-phonon coupling. Specifically, phonon excitations corresponding to displacements of the T layers along $\hat{z}$ couple to the electronic charge density of the adjacent H-layers, as shown in Fig.~\ref{Fig:4Hb_structure}(b). Accordingly, \hb is expected to host enhanced FE fluctuations. In this case, we consider the application of our theory to an ``antiferroelectric'' mode whereby opposite charge builds up on the two H layers. As we show below, this mode undergoes an anomalous softening due to the nonlinear capacitance of the system, making it an attractive candidate for FE pairing.

\subsubsection{Characterization of the Phonon}

As explained, 4Hb-TaS$_2$ has a quadruple layer unit cell and therefore has three nonequivalent optical modes associated with out-of-plane \emph{breathing} distortions.
We begin our analysis by determining, which of these is most suitable to be considered as the displacement $u$ in Eq.~\eqref{Eq:S_psi_u}. We focus on the region near the $\Gamma$-point of the Brillouin zone. The electronic bands of the two H-layers 
are approximated as simple parabolic bands, as described in Eq.~\eqref{Eq:S_psi}. 
This approximation, together with the omission of the Coulomb repulsion, renders the intra-layer pairing state a simple s-wave state.  Adding these components may induce non-s-wave pairing states, which will not be explored in this paper.

To characterize phonons polarized in the $\hat{z}$-direction, we consider a four-atom unit cell corresponding to the layers of H-T-H$^\prime$-T$^\prime$. In this approach, each layer is treated as a single ``effective atom", with motion restricted along $z$. This configuration leads to four phonon branches at each wave vector. For a qualitative understanding, we assume the layers are identical in mass. 
At the $\Gamma$-point, three optical modes emerge, including a pair of degenerate modes with displacement patterns schematically represented by
($\cdot$ $\rightarrow$ $\cdot$ $\leftarrow$ and $\rightarrow$ $\cdot$ $\leftarrow$ $\cdot$), where the arrows indicate the relative motion of each layer.

The first mode, as an example, can be understood via the distortion of the T and T$^\prime$ layers toward the H$^\prime$ layer, effectively causing charge transfer away from the H-layer to the H$^\prime$ layer. This results in an interlayer dipole, and induces a coupling proportional to the charge density difference between the H and H$^\prime$ layers [Fig.~\ref{Fig:4Hb_structure}(b)]. This effect 
 corresponds to the electron-phonon interaction term  
$S_{\psi u}$ in Eq.~\eqref{Eq:S_psi_u}.

The remaining modes, ($\rightarrow\,\cdot\,\leftarrow\,\cdot$) and the so-called ``dimerization'' mode
($\leftarrow\,\rightarrow\,\leftarrow\,\rightarrow$), asymmetrically displace the T-layers relative to the
H-layers. From the perspective of the H-layer, this generates an effective perpendicular electric field.
In the presence of spin--orbit coupling (SOC), these modes can induce Rashba-like terms of the form
$(\mathbf{k}\times\vec{s})\otimes\tau_z$ and $(\mathbf{k}\times\vec{s})\otimes\tau_0$, respectively.
However, we neglect their contributions hereafter, since the dominant Ising SOC splits the Fermi surfaces
and polarizes the spins along the $z$ direction, suppressing the effects of Rashba SOC
(see Sec.~\ref{sec:model}).

The phonon displacement patterns and their corresponding induced couplings are summarized in Table~\ref{Tab:phonon_modes}.
In accordance with Eq.~\eqref{Eq:S_u}, we treat the phonon field $u_{\qb}$ as a scalar, retaining only the out-of-plane displacement. As we noted above, this is an antiferroelectric mode a total FE polarization of zero even in the ordered state, but with a nonzero interlayer FE polarization.

\begin{table}[t]
\centering
\begin{tabular}{|c|c|}
\hline
\makecell{Optical mode in \\ H-T-H$^\prime$-T$^\prime$} & Induced coupling \\
\hline \hline
\( \cdot \ \rightarrow \ \cdot \ \leftarrow \) & $\sigma_0\otimes\tau_z$ \\
\hline
\( \rightarrow \ \cdot \ \leftarrow \ \cdot \) &  \( (\kv \times \vec{\sigma}) \otimes \tau_z \) \\
\hline
\( \leftarrow \ \rightarrow \leftarrow \ \rightarrow \) &  \( (\kv \times \vec{\sigma})\otimes \tau_0 \) \\
\hline
\end{tabular}
\caption{Optical phonon modes at the \( \Gamma \)-point and their associated couplings. Arrows indicate the direction of layer displacement.}
\label{Tab:phonon_modes}
\end{table}

\subsubsection{Softening of the phonon mode}
Above we identified the relevant breathing mode that appears in the coupling between charge and the anti-FE mode, Eq.~\eqref{Eq:S_psi_u}. 
We now show that charge transfer causes this  mode to become anomalously soft.
Indeed, the elastic energy cost associated with the displacement of the middle layer (T) is renormalized down by 
electrostatic interactions induced by charge transfer. 

The essential argument why the phonon is softened by charge transfer can be understood using an analogy to a parallel-plate capacitor. When the T-layer is displaced from its equilibrium position, the charge is redistributed to the adjacent H-layers. Assuming a uniform ($\qv=0$) displacement $u$ along $\hat{z}$, the charge densities in the top ($\rho_t$), and bottom ($\rho_b$), and middle ($\rho_T$) layers can be expanded as:
\begin{align}
 &\rho_t=\rho_0(1+c_1 u+ \mt{O}(u^2)+..)\,, \\
 &\rho_b=\rho_0(1- c_1 u+ \mt{O}(u^2)+..)\,, \\
 &\rho_T=\rho_0(1- \mt{O}(u^2)+..)\, ,
\end{align}
where $\rho_0$ is the equilibrium charge density and $c_1$ is just the charge transfer rate. 
The electric field generated between the top and bottom layers due to the charge imbalance depends linearly on $u$ at the leading order:
\begin{equation}
E_{tb}(u)=\dfrac{\rho_t-\rho_b}{2\ep}\approx \dfrac{c_1}{\ep} \rho_0 u +...\,,
\end{equation}
where $\ep$ is the dielectric constant. 
The corresponding electrostatic energy per unit cell area $\sim a^2$ is given by:
\begin{equation}
\mathcal{E}_c= -a^2 \int_0^u du' \rho_T E_{tb}(u') \approx -\left(\frac{c_1 a^2 \rho_0^2}{2\epsilon}\right)u^2\, .
\end{equation}
This energy gain represents the fact that due to the charge transfer, capacitative coupling drives the system to \emph{bigger} rather than smaller deformations.

Let us now produce an order-of-magnitude estimate for the softening. The calculations of Ref. \cite{Crippa_ncom2024} indicate that the equilibrium charge transfer is on order of one electron per star-of-David, i.e.  $\rho_0\sim e/(13 a^2)$, where $a\approx 3.4${\AA} is the \hb in-plane lattice constant . We estimate  
$c_1^{-1}\sim(3/2)${\AA}  
from the linear slope in Fig.~2(c) of that paper.
This produces $\mathcal{E}_c\sim -0.3 (u/a)^2$ eV, which represents a substantial energy scale, especially since this is an out-of-plane displacement. An estimate from the structural distortion itself yields 
$\langle u\rangle/a\sim 10^{-1}$~\cite{Xray_1T_Albert1997}, leading to a reduction in phonon energy by $\sim 3$ meV. Such a reduction is comparable with typical phonon energies and therefore we cannot rule out a significant softening of the mode and even a possible low-energy ordered state. This feature is unique to the 4 layer structure of \hb and establishes it as a promising candidate for (anti-)FE mediated SC. For this reason the plots of superconducting $T_c$ in Sec. \ref{sec:model} were computed using \hb parameters.

\subsubsection{Breaking of time-reversal symmetry}
Experimental muon spin relaxation measurements in \hb indicate that the superconducting (SC) state breaks
time-reversal symmetry $\mt{T}$~\cite{RibakChiralSC_SciAdv2020}. Moreover, the observation of magnetic memory---where
a magnetic field applied between $T_c$ and a higher onset temperature $T^*=3.6$~K modifies the SC state upon
cooling---suggests that $\mt{T}$ may already be broken in the normal state~\cite{persky2022magnetic}.
Consistent with these observations, the SC phase has been proposed to be chiral~\cite{Silber2024,Amit_NatCom2021}.

In Sec.~\ref{SSec:TRSbrokenSC} we identified the symmetry conditions under which the superconducting order parameter can couple
sympathetically to magnetism via Eq.~\eqref{eq:trs}: inversion symmetry and at least two mirror symmetries must
be broken. Notably, \hb lacks both $\hat{x}$ and $\hat{y}$ mirror symmetries, with the latter removed by the
SoD charge-density-wave phase. Consequently, if \hb additionally undergoes inversion symmetry breaking due to
charge-transfer–driven lattice softening (i.e., $u_0\neq 0$), the coupling in Eq.~\eqref{eq:trs} becomes symmetry
allowed. 
This provides a natural microscopic route for superconductivity in \hb to link
time-reversal symmetry breaking with superconductivity. It is also worth mentioning that the ratio between the coherence length and the mean-free path  puts \hb in the dirty limit. Thus, the scenario of an $s+i\,s$ order parameter is advantageous over those invoking a non-s-wave chiral order parameters due to its robustness to disorder.

\subsection{Interlayer shear modes -- application to sliding ferroelectrics} 

Sliding FE materials are a family of bilayer or few-layer vdW systems where out-of-plane inversion is broken via a sliding motion of the layers (\emph{shear} motion). In sliding FEs, each layer typically contains at least two nonequivalent atoms per unit cell. A small in-plane displacement between adjacent layers alters the stacking configuration, leading to a change in the local dipole moment. This change effectively induces a local charge redistribution between the layers. Such sliding instabilities have been found in both insulating and metallic materials. In addition, because of the weak vdW forces between the layers, the vibrational frequencies corresponding to the sliding motion tend to be quite soft. This places sliding FE metals as another promising candidate for FE superconductivity.

\subsubsection{Construction of the model}


\begin{figure*}
    \centering
    \includegraphics[width=\linewidth]{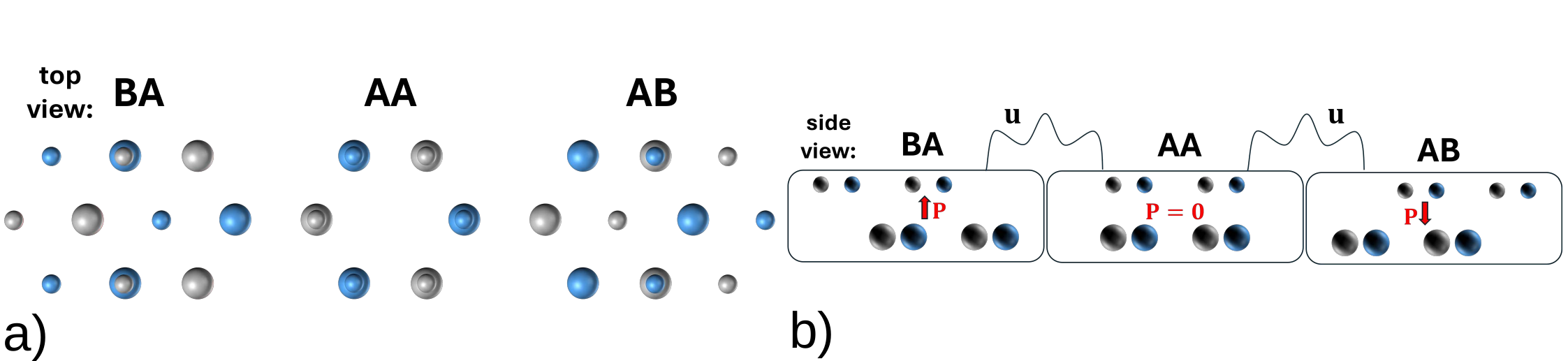}
    \caption{Illustration of the sliding procedure. (a) A top view of a hexagonal bilayer, with an unstable inversion-even AA stacking separating two inversion-broken configurations. The atoms in the two layers, although physically the same, have been rendered in different sizes for visual clarity. (b) An illustration of how the inequivalent atoms give rise to an interlayer polarization. The transition between configurations is mediated by the interlayer shear mode that we model as $u$.}
    \label{fig:slideSchem}
\end{figure*}

To construct our theory, consider a simple model of a two component hexagonal lattice, stacked into a bilayer (Fig. \ref{fig:slideSchem}), which describes e.g. sliding hexagonal Boron Nitride (hBN). This gives rise to a doubly degenerate low energy polarized configuration (AB/BA stacking) separated by a higher energy nonpolar one (the AA stacking). The ``sliding'' motion from the low-energy to high-energy configuration is just an interlayer shear mode (TA' mode), which is acoustic in the plane an optical out-of-plane. \AKn{In practice, the unstable configuration is often not AA  but some other inversion-even state with lower energy, but this does not affect the form of the resulting model (see more below).} Similarly to what occurs in bulk FE materials, one expects the in-plane component to be transverse so as to minimize the formation of bound charge. 

Neglecting for simplicity the threefold rotational symmetry of each layer, the energy landscape of the transition is a double well, with AB/BA global minima and AA a local maximum. Despite the fact that the transition temperature is typically above room temperature, studies indicate that the system is well described by a standard Landau-type quartic theory \cite{Yang2018,PhysRevLett.130.176801}. In addition, the relevant shear modes are extremely soft, on order of only 1-2 meV. The implication is that a convenient starting point for a theory of sliding FEs is not the polarized ground state but rather the unstable, nonpolar one. The order parameter of the transition is then nothing but the amplitude of the shear phonon, which breaks inversion symmetry. We then immediately obtain Eq. \eqref{Eq:S_u}, with the caveat that we should perform all computations in the ordered state. The generalization to a metallic bilayer is immediate, giving rise to precisely the model of Sec. \ref{sec:model}. \AKn{Note, that the coupling term, Eq. \eqref{Eq:S_psi_u}, assumes implicitly a direct coupling between the in-plane mode and out-of-plane deformation. In the presence of rotational symmetry, such coupling is often forbidden and hence some mechanism to break it, e.g. via strain or a lattice broken in-plane mirror, is necessary to provide the coupling. This will induce some angular form-factors, which we neglect for simplicity in this work.}

\subsubsection{Application to MoTe$_2$}

MoTe$_2$ bilayers have recently been suggested as condidates for superconducting sliding FEs. The SC temperature in MoTe$_2$ is strongly layer dependent and increases down to the bilayer, suggesting that screening effects are crucial. In addition, SC $T_c$ increases dramatically upon application of an electric field opposite to the spontaneous polarization. We now show that our model is a reasonable candidate to describe MoTe$_2$.

MoTe$_2$ is a nearly compensated metal, but it has two reasonably large FSs around the $\Gamma$ point. The larger one is well approximated as a parabolic hole band. Its Curie temperature to the polarized state is approximately $350$K \cite{Yuan2019RoomTemperature}, but the energy of its interlayer shear mode is significantly lower \cite{Fukuda2020}. 

At low temperature, as the system approaches the FE quantum critical point—where the softening of the sliding phonon modes becomes significant—superconductivity can emerge with the same symmetry properties as discussed in Secs.~\ref{Sec:pair_sym} and \ref{Sec:sol_gap}.  Some estimates of the relevant parameters are presented in Table \ref{tab:mote2-params}. From these we may estimate the validity of the action in Eqs. \eqref{Eq:S_u}-\eqref{Eq:S_psi_u}. A quick calculation yields
\begin{equation}
    r \approx 0.2,\,  k_F\xi \approx 3.5\, ,
\end{equation}
implying this system is reasonably well approximated by the quantum-critical theory we presented. Here we recall that $r$ is the distance from the critical point, Eq.~\eqref{eq:rdef}, and $\xi = 2\pi a/\sqrt{r}$ is the correlation length. Furthermore, from the tight-binding model presented in Ref. \cite{Jindal2023}, we may estimate the the splitting due to the ordered state is of order $\delta\mu \sim 20 \mbox{meV}$. Such an estimate yields a $\bar{\lambda} = \delta\mu/u_0$ which even for $u_0 = O(1)$ in the ordered state, gives rise to $T_c$ values that are in quite reasonable agreement with the experiment, see further discussion below and in Appendix \ref{app:MoTe2}.

We now show that, similarly to \hb, sFE SC may be strongly affected by capacitative effects. One of the puzzles raised by Ref. \cite{Jindal2023} is the strong enhancement (approximately $\times 2$) of the SC $T_c$ upon application of an out-of-plane electric field, but only when it is \emph{opposed} to the spontaneous polarization. In fact, $T_c$ went up until the applied field reached the coercive field strength, at which point the polarization reversed and SC disappeared. Such behavior is strange since the electric field $E$, the polarization $P$ and the order parameter $u_0$ all break inversion symmetry explicitly, which does not increase pairing fluctuations and thus naively should not enhance $T_c$. Within our model, however, a coupling term arises that naturally explains the effect, in the form of the capacitative energy. To see this, consider the following expression for the field-induced polarization energy. It is nothing but the parallel-plate capacitor energy given the interlayer distance,
\begin{align}
    \label{eq:sFE-capacitance}
    \mathcal E_P = -EP &= -E (-e\Delta n(\mathbf{x})) d(\mathbf x) \nonumber \\
    &= -E\left(-e\bar{\psi}(\mathbf{x})\tau_z\psi(\mathbf{x})\right)d_0(1-\alpha_0 u^2(\mathbf{x}))+\cdots\, ,
\end{align}
where $d$ is the interlayer distance, $\Delta n(\mathbf{x})$ is the electron transfer between layers which is negative for a positive field, and $\mathbf{x}$ is an in-plane coordinate. The quadratic correction accounts for the fact that upon sliding the layers will move closer together so as to reduce their intrinsic (not field induced) capacitative energy, where $\alpha_0$ is a geometric factor, such that 
\[d(\mathbf x)\approx d_0 (1-\alpha_0 u^2(\mathbf{x})+\mathcal O(u^4))\, .\]
Eq. \eqref{eq:sFE-capacitance} represents a quartic (in terms of inversion-breaking fields) term in the free energy. However, in the ordered state \cite{PhysRevB.106.L121114} such contributions directly impact the linear coupling. Indeed, upon linearization it yields a term~\cite{Saha2025Strong},
\begin{equation}
    \label{eq:cap-coupling-correction}
    \delta H_I = -2\alpha_0Eedu_0\sum_{\kv,\qv}\bar{\psi}_k\tau_z\psi_{k+q}\delta u_q+\cdots.
\end{equation}
Thus, the term directly enhances the pairing coupling but only when the field is opposite to the spontaneous order, $E u_0 < 0$. Since the only nonuniversal term in the prefactor is the order parameter itself, one expects this enhancement to be significant. Further, it vanishes precisely at the coercive field, since the moment $u_0$ switches sign the correction is an effective reduction.

We do however note one caveat. While we were not able to find estimates of the equilibrium shift of the sliding motion, estimates for the structurally similar WTe$_2$ are in the order of $a_0 \sim 0.25$\AA. In terms of our theory, this give $u_0 \sim a_0/\ell_{osc} \sim 4$, where $\ell_{osc}=\sqrt{\hbar^2/2M m_r}$
is an extremely naive estimate for the oscillator length and $M$ is the atomic mass. Thus, the assumption of degenerate channels may be irrelevant for MoTe$_2$, reducing the chance for TRSB. Interestingly, taking the $u_0=4$ value along with the bulk phonon energy for $D_0 \sim a^2m_r^{-1}$, gives (along with the other parameters discussed above) a value of $\bar{\lambda}$ that is in semi-quantitative agreement with experiment. We note that numerical studies have shown that the system can be driven even closer to the critical point by strain, suggesting a promising route to drive TRSB in these systems.

\begin{table}[t]
    \begin{tabular}[t]{|c|c|c|}
        \hline
        Parameter & Estimate & Source \\
        \hline\hline
        $k_F$ & 0.07\AA$^{-1}$ & Theory \cite{Jindal2023}\\
        $m_r$ & 1.6 meV & Exp (bulk) \cite{Fukuda2020} \\
        $c$ & 2000 m/s & Theory \cite{Rano2020} \\
        $a$ & 3.5\AA & Exp \cite{Puotinen1961} \\
        \hline
    \end{tabular}
    \caption{Estimates of parameters for an MoTe$_2$ sliding FE system.}
    \label{tab:mote2-params}
\end{table}

\section{Discussion}
In this work, we developed a minimal theoretical framework for superconductivity mediated by interlayer ferroelectric charge-transfer fluctuations in van der Waals heterostructures. We showed that soft inversion-breaking interlayer phonon modes, which are often regarded as weakly coupled due to large layer separations, can instead provide a strong pairing channel once charge transfer between layers is properly taken into account. Near a ferroelectric or antiferroelectric quantum critical point, these modes become strongly renormalized and generate an effective attractive interaction that favors layer-triplet superconducting pairing. Within an SU(2)-symmetric framework, the resulting pairing state is intralayer and s-wave in the plane. We identified an accidental degeneracy between distinct layer-triplet pairing channels, which persists even when the in-plane gap structure is fully isotropic.

A key consequence of this quasi-degeneracy is the possible emergence of TRSB superconductivity. We showed, that the presence of static or even fluctuating  magnetic correlations will in general favor a TRSB interlayer triplet state. We found a concrete mechanism in the form of
interlayer Josephson coupling, where symmetry-breaking interlayer tunneling can select a complex relative phase between the superconducting order parameters of different layers. Such a mechanism may produce circulating interlayer supercurrents and spontaneous magnetic fields, providing a direct route to experimentally observable signatures of unconventional superconductivity. 
We applied our theory to two experimentally relevant platforms—4Hb-TaS$_2$ and sliding ferroelectric metals such as bilayer MoTe$_2$. We showed that charge-transfer-induced softening of interlayer modes can realistically account for observed superconducting energy scales and symmetry-breaking phenomena. 
Thus,  our results emphasize the importance of ferroelectric instabilities in the presence of charge transfer in the context of 4Hb-TaS$_2$~\cite{Kanigel_npj2024} and related materials~\cite{almoalem2025mixed}, whose fascinating charge dynamics tend to be eclipsed by other effects, e.g. TRSB, SC and CDW physics.

It is worth discussing the specific case of \hb a bit further, since, in contrast to e.g. the sliding MoTe$_2$ case, this platform has been studied extensively in recent years. We focused on the antiferroelectric distortion corresponding to a symmetry-broken stacking pattern of $T\to H' \leftarrow T'$ followed by an isolated $H$ layer. However, other symmetry-breaking distortions such as dimerization of the $T\to H$ and $T'\to H'$ layers are also possible. This deformation also breaks inversion symmetry and 
may be relevant in various experimental realizations.

As regards the possible TRSB in \hb, within the framework we have put forward it is intimately tied to the (anti-)ferroelectric (FE) order via the pseudovector $\mathcal{K}$ defined in Sec.~\ref{Sec:Interlayerhopping}. Experimentally, TRSB persists up to a temperature $T^* = 3.6\,\mathrm{K} \gtrsim T_c$. Within our model, $T^*$ may be associated either with an intrinsic magnetic instability tuned by the control parameter $r$ toward a critical point at $r=0$, or with a precursor phase driven by superconducting fluctuations through Eq.~\eqref{eq:trs}. 

Further insight into the nature of this phase is provided by recent measurements on 4Hb-TaSSe~\cite{Guo2025}, which demonstrate that Se substitution drives the $T$ layers metallic and eventually superconducting. We therefore expect Se doping to influence the FE instability in two distinct ways. At low concentrations, Se substitution acts as a source of quenched \emph{random-field disorder}: owing to the larger ionic radius of Se compared to S, local layer expansion favors one of the two inversion-related FE configurations, thereby coupling linearly to the discrete (Ising-like) FE order parameter and explicitly breaking the symmetry between the two configurations.
Given the quasi-two-dimensional nature of the FE order and the weak interlayer coupling in the 4Hb structure, the relevant Imry--Ma physics is effectively two-dimensional. In this case, arbitrarily weak random-field disorder is sufficient to destroy true long-range FE order, leading instead to the formation of finite FE domains. As a consequence, while local TRSB may persist within individual domains, global phase-coherent TRSB is expected to be suppressed already at small Se concentrations.

At higher Se concentrations, the $T$ layers become metallic, which is expected to further destabilize the interlayer FE order through enhanced electronic screening. We therefore \emph{conjecture} that global TRSB coherence is ultimately lost in the vicinity of the insulator--metal transition of the $T$ layers. A crude percolative argument suggests that this loss of coherence may occur once  half the stars-of-David are locally perturbed by Se substitution (considering site percolation on a triangular lattice). Interpreting this threshold at the level of order unity, one is led to an estimated Se concentration as low as $x \sim 0.02$, given that each star-of-David is surrounded by 52 S atoms. While this estimate is necessarily approximate and geometry dependent, it serves to highlight the extreme sensitivity of the FE order—and the associated TRSB—to disorder in this system.
 We note that there are experimental indications that magnetic memory in \hb is indeed destroyed at about that doping \footnote{B. Kalisky and A. Kanigel, private communication}. Of course, far more work is necessary to confirm or dispel this conjecture.


In this paper, we considered a simplified model with SU(2) symmetry and neglected Coulomb repulsion, which naturally leads to an s-wave pairing state. In more realistic settings, these assumptions can be relaxed and may give rise to genuinely unconventional pairing states. Such possibilities are particularly intriguing in the context of 4Hb-TaS$_2$, where signatures of nematic superconductivity have been observed~\cite{Silber2024,Amit_NatCom2021}. For example, our mechanism naturally allows for superconducting gaps that deform under strain while simultaneously breaking time-reversal symmetry, providing a promising direction for future theoretical and experimental exploration.
\AK{Considering our work from a bird's-eye perspective, it serves to point out the till-now somewhat ignored importance of charge transfer as a \emph{fluctuating}, dynamical process. The number of candidates for TRSB-SC that have an inherently layered structure is quite significant, and we expect this to be a promising future study avenue. We also hope our work will bring yet more attention to the fascinating topic of sliding TMDs and specifically to its metallic sub-family. One important property neglected in the current study is the intimate connection between the sliding dynamics and the motion of domain walls \cite{Chaudhary2024,Kopasov2024,Naumis2023,Bian2024}, which renders a mean-field treatment somewhat naive and may dramatically affect SC properties.}

\acknowledgements
We thank A. Chubukov, R.M. Fernandes, J. Schmalian, D. Pelc, H. Beidenkopf, B. Kalisky, A. Kanigel, E. Persky, and A. Almoalem  for many helpful discussions, and the authors of Ref. \cite{Jindal2023} for sharing their data with us. A. Klein acknowledges the support of the Israeli Science Foundation grant no. 3152/25. JR acknowledges funding by the Simons foundation and the ISF under grant No. 915/24.

\onecolumngrid
\appendix

\section{\label{appA}Self-energies and Pairing Equation in the Disordered State}

In what follows, we present the technical steps leading to the effective pairing 
discussed in the main text. We begin from a two-dimensional fermion-boson model in which 
out-of-plane FE phonons couple to the electronic layer degree of freedom. The 
relevant electron--phonon interaction, given in Eq.~\eqref{Eq:S_psi_u} of the main text, is 
reproduced here for completeness: 

\begin{equation}
S_{\psi u}=\lmb \Tr \sum_{\kb,\qb} u_{\qb} \bar{\psi}_{\kb} \tau_z\s_0 \psi_{\kb+\qb}\,,
\label{Eq:App_SpsiU}
\end{equation}

Here $\sigma_i$ denote Pauli matrices in spin space and $\tau_i$ denote Pauli matrices in the 
two-layer (pseudospin) space. The vertex $\tau_z\sigma_0$ represents a spin-independent 
coupling ($\sigma_0$) to the electronic density imbalance between the two layers ($\tau_z$). 

\subsection{Self-energies}
We start with the normal-state bosonic and fermionic self-energies, which are computed to one-loop order. 
The bosonic self-energy [Eq.~\eqref{Eq:Pi}], obtained from the fermionic bubble shown diagrammatically in 
Fig.~\ref{Fig:F_diagram}(a), can be written as


\begin{equation}
 \begin{aligned}
\Pi(\bar{q}) &=\frac{\bar{\lambda}}{k_a^2}\, T \, \sum_{k_{0}} \int \frac{d^2\bm{k}}{\left(2\pi\right)^2}\text{Tr}\left[\tau^z \sigma_0 G(\bar{k}) \tau^z \sigma_0 G(\bar{k}+\bar{q})\right]  \\
&=4\frac{\bar{\lambda}\nu_F}{2\pi} \int \left(\frac{1}{(k_0+q_0)+i\epsilon_{\bm{k}+\bm{q}}}\right) \left( \frac{1}{k_0+i\epsilon_{\bm{k}}} \right)  \,   \frac{dk_0}{2\pi} \,  d\epsilon_{\bm{k}} \, d\theta    \\
&=4\frac{\bar{\lambda}\nu_F}{2\pi}  \int \frac{2\pi i \left[\theta(-\epsilon_{\bm{k}})-\theta(-\epsilon_{\bm{k}+\bm{q}})\right]}{2\pi \left(i v_F q \, \hat{k}\cdot\hat{q}   + q_0 \right)}    \, d\epsilon_{\bm{k}} \, d\theta    \\
&=4\bar{\lambda}\nu_F\left[1-  \frac{\vert q_0 \vert}{\sqrt{q_0^2+\left( v_F q\right)^2}}  \right] \, .
\label{eq:bubble_2D_full}
 \end{aligned}
 \end{equation}
Here, $\nu_F=m^{*} a^2/(2\pi)=k_F a^2/2\pi v_F$ is the 2D density of states (DOS) per spin at the Fermi level and $k_a=1/a$ is the inverse lattice parameter, $q=|\qv|$.
The trace in the first line of  Eq.~\eqref{eq:bubble_2D_full} runs over spin and layer indices, and yields an overall factor of 4. 
%
In the limit $\vert q_0 \vert \ll  v_F \vert \bm{q} \vert$, the polarization bubble in Eq.~\eqref{eq:bubble_2D_full} can be approximated as
\begin{equation}
 \begin{aligned}
\Pi(\qb) &\approx 4\bar{\lambda}\nu_F \left( 1-   \frac{\vert q_0 \vert}{v_F q} \right).
 \label{eq:bubble_2D_LD}
  \end{aligned}
 \end{equation}
as shown in Eq.~\eqref{eq:bubble_2D_LD1} in the main text. 
Next, we focus on the fermionic self-energy [Eq.~\eqref{Eq:Sigma}] in the disordered state. It is shown diagrammatically in Fig. \ref{Fig:F_diagram}(b). It assumes the following form
 \begin{equation}
 \begin{aligned}
         \Sigma(k_0) &=\frac{\bar{\lambda}}{ k_a^2}   T \sum_{q_0}  \int \frac{d^2q}{(2\pi)^2}\,\tau^z \sigma_0 G(\bar{k}+\bar{q}) \tau^z \sigma_0 D(\bar{q})\, , \\
         &=\frac{\bar{\lambda}}{ k_a^2} \frac{1}{(2\pi)^2}  T \sum_{q_0}  \int q D(\bar{q}) \left[\int d\theta \frac{1}{i(k_0+q_0)-v_F q \cos\theta}   \right]dq\, \\ 
         &\approx \frac{-i\bar{\lambda}\nu_F}{ k_Fa}  T \sum_{q_0} \text{sgn}(k_0+q_0) d(q_0),
          \label{eq:fermion_self_energy}
 \end{aligned}
\end{equation}
where 
the frequency-dependent function $d(q_0)$ is obtained by integrating out the momentum in the bosonic propagator $D(\bar q)$,  
 \begin{equation}
 \AK{d(q_0)=\frac{1}{k_a}\int_0^{\Lambda} dq  \left(   \frac{1}{r+\frac{q_0^2a^2}{c^2}+q^2 a^2+4\bar{\lambda} \nu_F \frac{\vert q_0 \vert}{v_F q}} \right)} \,,
          \label{eq:boson_function_full2}
\end{equation}
where we have used the expression for the polarization bubble in Eq. \eqref{eq:bubble_2D_LD}. The distance from QCP is measured by $r$, which is defined in the main text as Eq. \eqref{eq:rdef}. In Eq. \eqref{eq:boson_function_full2}, $\Lambda$ is an UV momentum cutoff which can be safely taken to infinity. For simplicity, we neglect the bosonic frequency term $(q_0 a/c)^{2}$. At
the QCP ($r = 0$) and in the limit $v_{F} q \gg q_0$, this bosonic function acquires the form 
\begin{equation}
d(q_0)=\frac{1}{k_a}\int_0^{\infty} dq \ \dfrac{1}{ q^2 a^2+\gamma \frac{|q_0|}{q}} 
= 
\frac{2\pi}{3\sqrt{3}}\Big(\dfrac{1}{a\, \gamma}\Big)^{1/3} |q_0|^{-1/3}\, .
\label{eq:identity}
\end{equation}
where $\gamma = 4 \bar{\lambda}\frac{\nu_{F}}{v_{F}}$.
Consequently, the fermionic
self-energy at the QCP scales as from Eq.\eqref{eq:fermion_self_energy},
\begin{equation}
\Sigma(k_0)
=
-i \omega_0^{1/3}
\mathrm{sgn}(k_0)\int_0^{|k_0|} d q_0  |q_0|^{-1/3} 
=
 -i \, \omega_0^{1/3}
 \mathrm{sgn}(k_0) 
 |k_0|^{2/3}\, , \omega_0 = \frac{\pi \bar{\lambda}^2}{48 \sqrt{3}\, k_F\, v_F }\, .
 \label{eq:k23}
\end{equation}
Thus, $\Sigma(k_0) \propto |k_0|^{2/3}$, signaling the breakdown of Fermi-liquid theory and the emergence of non-Fermi-liquid behavior driven by critical bosonic fluctuations.

\subsection{Gap equation}
Now, we turn our attention to the linearized equation for the pairing vertex, shown diagrammatically in Fig. \ref{Fig:F_diagram}(c). It assumes the form

\begin{equation}
  \Phi(k_0) =\frac{\bar{\lambda} T}{ k_a^2} \sum_{q_0} \int \frac{d^2q}{(2\pi)^2} \tau_z\s_0  G(\bar{k}+\bar{q}) D(\bar{q}) G(-\bar{k}-\bar{q}) \Phi(k_0+q_0) \tau_z\s_0\, .
  \label{eq:phidis}
\end{equation}


Where, the structure of the interaction vertex $\tau_z$ restricts the pairing kernel to the 
$\tau_0$ and $\tau_z$ components in layer pseudospin space. This follows from the relation
$\tau_z \tau_i \tau_z = \eta_i \tau_i $,
$\eta_i = \{+1,-1,-1,+1\} $\; for $ i=\{0,x,y,z\}.
$
Hence the attractive pairing channels are spin singlets of the form 
$i\sigma_y \tau_0$ and $i\sigma_y \tau_z$.
We can simplify Eq. \eqref{eq:phidis} as

\begin{equation}
 \begin{aligned}
	 \Phi(k_0) &=\frac{\bar{\lambda}}{ k_a^2}  \frac{1}{(2\pi)^2}  T \sum_{q_0} \int q  D(\bar{q}) \Phi(k_0+q_0) \left[\int_0^{2\pi} \frac{d\theta}{\vert \tilde{\Sigma}(k_0+q_0) \vert^2+\epsilon_{\bm{k}+\bm{q}}^2}   \right] dq \\
	 &=\frac{\bar{\lambda}}{ k_a^2}  \frac{1}{(2\pi)^2}  T \sum_{q_0} \int q  D(\bar{q}) \Phi(k_0+q_0) \left[\int_0^{2\pi} \frac{d\theta}{\vert \tilde{\Sigma}(k_0+q_0) \vert^2+(v_F q)^2 \cos^2\theta}   \right] dq \\
     &=\frac{\bar{\lambda}}{ k_a^2}  \frac{1}{2\pi}  T \sum_{q_0} \int q  D(\bar{q}) \Phi(k_0+q_0) \frac{1}{(v_F q)^2} \left[    \frac{1}{  \frac{\vert \tilde{\Sigma}(k_0+q_0) \vert} {v_F q } \sqrt{1+\frac{\vert \tilde{\Sigma}(k_0+q_0) \vert^2} {(v_F q)^2}}}   \right] dq \\
	 &\approx\frac{\bar{\lambda}\nu_F}{ k_Fa}  T \sum_{q_0} \frac{\Phi(k_0+q_0) d(q_0)}{\vert \tilde{\Sigma}(k_0+q_0) \vert} \, . 
 \label{eq:vertex}
 \end{aligned}
\end{equation}

where we have defined 
$\tilde{\Sigma}(k_0) = k_0 + \Sigma(k_0)$. In the last step of Eq. \eqref{eq:vertex}, we have used the limit $\tilde{\Sigma}(k_0+q_0) \vert \ll  v_F \vert q \vert$.

\subsection{Exclusion of thermal contribution from the gap equation in disordered state}

Both the expression for self-energy $\Sigma(k_0)$ (Eq. \eqref{eq:fermion_self_energy}) and the pairing vertex $\Phi(k_0)$ (Eq. \eqref{eq:vertex}) contain 
contributions from the $q_0=0$ Matsubara mode, which diverge at the (QCP). This is formally analogous to the effect of
non-magnetic impurity scattering and, in accordance with Anderson’s theorem Ref \cite{Anderson1959DirtySuperconductors}, does not affect $T_c$. There is a well known procedure to discard this static contribution in the disordered state, since it does not affect superconducting instabilities.
Following standard practice, 
we exclude this static contribution from the Matsubara summation and retain only the 
dynamical sector ($q_0 \neq 0$), by introducing a new gap function $\Delta(k_0)=k_0\,\Phi(k_0)/\tilde{\Sigma}(k_0)$ 
(See Ref. \cite{MillisSachdevVarma1988Inelastic, AbanovChubukovNorman2008GapAnisotropy, MoonChubukov2010QuantumCritical, ChubukovAbanovWangWu2020InterplaySuperconductivity} for a detailed discussion). The pairing function in Eq. \eqref{eq:vertex} now reduces to the following eigenvalue problem. 

\begin{equation}
\Delta(k_0)=
\frac{\displaystyle 
\frac{\bar{\lambda}\,\nu_{F}}{k_{F}a}\;
T \sum_{p_0 \ne k_0} 
\frac{1}{p_0}\bigl[d(p_0-k_0)+d(p_0+k_0)\bigr]\,
\Delta(p_0)}
{\displaystyle 
1+\frac{\bar{\lambda}\,\nu_{F}}{k_{F}a}\;
T \,\frac{1}{k_0}
\sum_{p_0^\prime \ne k_0}
\bigl[d(p_0^\prime -k_0)-d(p_0^\prime +k_0)\bigr]}\, .
\label{eq:eliashberg_gapeq_matrix}
\end{equation}
Here, the UV cutoff in the frequency summation is the Fermi energy $E_F/k_B$, $k_B$ is the Boltzmann constant. In Fig. \ref{Fig:Tc_coupling} we take the Fermi momentum $k_F = 0.59~\mathrm{\AA^{-1}}$ from ARPES measurements \cite{Kanigel_npj2024} on \hb. Solving Eq.~\eqref{eq:eliashberg_gapeq_matrix} numerically, we identify $T_c$ as 
the highest temperature at which a nontrivial solution for $\Delta(k_0)$ exists, corresponding to the point where the largest eigenvalue of the kernel 
reaches unity (see green line in Fig.~\ref{Fig:Tc_coupling}).

\section{\label{app:ordered}Self-energies and pairing equation in the ordered state}

Here we extend the theory of the main text to the ordered state ($r<0$). We adopt a
Ginzburg--Landau description for the bosonic sector with the order parameter $u_{\bar{q}}$.
The free energy (up to quartic order) is taken as
\begin{equation}
F = -|r|u_{\bar{q}}^2 + \frac{b}{2}u_{\bar{q}}^4 \,, \qquad
\label{eq:free}
\end{equation}
where $r$ is the quadratic Ginzburg–Landau coefficient that tunes the instability between ordered and disordered state, and $b>0$ is the quartic coefficient which stabilizes the free energy in the ordered state. The free energy in Eq. \eqref{eq:free} is minimized by $u^2_0=r/b$. We therefore write the displacement field as $u_{\bar{q}}=u_0+\delta u_{\bar{q}}$, where $\delta u_{\bar{q}}$ denotes small fluctuations around $u_0$. Rewriting the bosonic action
[Eq.~\eqref{Eq:S_u}] in terms of $\delta u_{\bar{q}}$ gives
\begin{equation}
S_u \;=\; -\left.\frac{r^2}{2b}\right|_{u=u_0} \;+\; \sum_{\bar q} \delta u_{\bar q}\,
\bigl[D^{-1}(\bar q)\bigr]\,\delta u_{-\bar q}\,,
\label{eq:suordered1}
\end{equation}
where the first and second terms describe the static and dynamic contributions (we write down the dynamical propagator explicitly below). The electron-phonon coupling [Eq.~\eqref{Eq:S_psi_u}] accordingly becomes
\begin{equation}
S_{\psi u} \;=\; \lambda \Tr \sum_{\bar{k},\bar{q}}\bigl(u_0 + \delta u_{\bar{q}}\bigr)\,
\bar{\psi}_{\bar{k}}\,\tau_z\sigma_0\,\psi_{\bar{k}+\bar{q}}\, ,
\label{eq:s_phiu_ordered}
\end{equation}
so the static component $u_0$ of the FE displacement acts as a 
layer-odd potential, shifting the electronic energies of the two layers, raising one and lowering the other. Consequently, the electronic part of the action $S_\psi$ [Eq.~\eqref{Eq:S_psi}] is modified to

\begin{equation}
S_{\psi} = \Tr \sum_{\bar k} \bar{\psi}_{\bar k}
\Bigl(i k_{0}-\frac{\mathbf{k}^2}{2m^*} +\mu \Bigr)\tau_0\sigma_0\,\psi_{\bar k}
+ \lambda u_0 \Tr \sum_{\bar k}\bar{\psi}_{\bar k}\,\tau_z\sigma_0\,\psi_{\bar k}\,,
\label{eq:spsiordered}
\end{equation}

which generates a layer-dependent chemical potential, where $\mu_t$ ($\mu_b$) corresponds to the top (bottom) layers.

\begin{equation}
\mu_{t}\,, \mu_b = \mu \pm \delta\mu\,, \qquad \delta\mu = \lambda u_0\,,
\end{equation}
as shown schematically in Fig.~\ref{Fig:Tc_coupling} leads to a splitting of the $T_c$ of the two layers. Equation~\eqref{eq:spsiordered} can be rewritten as
\begin{equation}
S_{\psi} \;=\; \sum_{\bar k} \bar{\psi}_{\bar k}\,
\bigl[G(\kb)\bigr]^{-1}\sigma_0\,\psi_{\bar k}\,,
\end{equation}
where the fermionic propagators of the top and bottom layers differ; the ordered-state Green’s function is
\begin{equation}
\begin{bmatrix}
  G_{t}(\kb) & 0\\
  0 & G_{b}(\kb)
\end{bmatrix}
 \;=\;
\begin{bmatrix}
\dfrac{1}{i k_{0} - \epsilon_k + \delta\mu} & 0 \\
0 & \dfrac{1}{i k_{0} - \epsilon_k - \delta\mu}
\end{bmatrix} \,, 
\label{eq:gtb}
\end{equation}
where $\epsilon_k=\mathbf{k}^2/(2m^*)-\mu$. From here on, we will use the subscript $l=t,b$ to denote the layer index for convenience.

\vspace{.5cm}

\subsection{Self energies: modifications relative to the disordered case}

Here we summarize the ordered state expressions obtained by substituting the ordered state propagators Eq.~\eqref{eq:gtb} into the one loop formulas derived in the disordered state, without repeating intermediate steps. Starting from Eq.~\eqref{eq:bubble_2D_full}, the bosonic self-energy in the ordered state reads

\begin{equation}
\begin{aligned}
\Pi(\qb) &= -2\bar{\lambda} T 
 \sum_{\kb; \,l = t,b}  G_l(\kb) G_l(\kb + \qb)\\
&\approx 4 \bar{\lambda} \nu_F 
\left(\AK{1}- \frac{1}{v_{eff}}\frac{|q_0|}{q}\right)\, ;\qquad \frac{1}{v_{eff}}= \frac{1}{2}\sum_{l} \frac{1}{v_{F l}}\, .
\end{aligned}
\label{eq:pi_order}
\end{equation}

Valid in the regime $|q_0|\ll v_{F l}|\mathbf{q}|$. Here $v_{F l}$  is the Fermi velocity of the top (bottom) layers, and same two-dimensional density of states per spin at the Fermi level as in the disordered state is $\nu_F$.
Using Eq.~(\ref{eq:pi_order}), the dressed bosonic propagator becomes\
\begin{equation}
D(\bar{q})=\frac{1}{\;2|r|+q^{2} a^2+q_0^{2}a^{2}/c^{2}
+4\bar{\lambda}\nu_F \frac{\vert q_0 \vert}{v_{eff} \,q}}\,.
%
\label{eq:Dorder}
\end{equation}
In the disordered phase, $r>0$ determines the bosonic propagator,
$D^{-1}(\bar{q}=0)\sim r$ (Eq. \eqref{eq:boson_function_full2}), while in the ordered state the curvature of the free energy Eq. \eqref{eq:free} at $u_0$ gives $D^{-1}(\bar{q}=0)\sim 2|r|$ (Eq. \eqref{eq:Dorder}) as the effective mass term.




From Eq.~\eqref{eq:fermion_self_energy}, the ordered-state fermionic self-energy becomes,

\begin{equation}
\Sigma =
\begin{pmatrix}
\Sigma_t & 0 \\[2pt] 0 & \Sigma_b
\end{pmatrix}, \qquad
\Sigma_{l}(k_0)
= \frac{-i\,\bar{\lambda}\,\nu_{F}}{k_{Fl}a}\;
T \sum_{q_0}\! \mathrm{sgn}(k_0+q_0)\, d(q_0)\,,
\label{eq:ferminicslfenergy_ordered}
\end{equation}

where the frequency-only kernel $d(q_0)$ is the same object defined in the disordered state
[Eq.~(\ref{eq:boson_function_full2})], now evaluated with ordered-state parameters:

\begin{equation}
d(q_0) = \frac{1}{k_a}\int_{0}^{\Lambda}
\frac{dq}{2|r| + q^{2} a^2
+4\bar{\lambda}\nu_F \frac{\vert q_0 \vert}{v_{eff} \,q}}\,.
\label{eq:dq0_integral}
\end{equation}

As in the disordered case, we neglect the term $(q_0 a/c)^2$.
At the QCP ($r\to 0$) and for $v_{Fl}|\mathbf{q}|\gg |q_0|$,
Eq.~(\ref{eq:dq0_integral}) yields $d(q_0)\propto |q_0|^{-1/3}$ (Eq.~\eqref{eq:identity}),
and consequently $\Sigma_{l}(k_0)\propto |k_0|^{2/3}$ (Eq.~\eqref{eq:k23}).

\subsection{Gap equation in the ordered state}

Similarly to the disordered state, Eq. \eqref{eq:vertex}, we can rewrite the linearized equation for the pairing vertex in terms of a layer-dependent term:
\begin{equation}
    \Phi_l (k_0) \approx \frac{\bar{\lambda}\nu_F}{ k_{Fl}a}  T \sum_{q_0} \frac{\Phi_l(k_0+q_0) d(q_0)}{\vert \tilde{\Sigma}_l(k_0+q_0) \vert}\,.
    \label{eq:gapEqOrdered}
\end{equation}

\subsection{\label{appA3} Exclusion of thermal contribution from the gap equation in the ordered state}
Similar to disordered state both the expression for self-energy Eq. \eqref{eq:ferminicslfenergy_ordered}) \AK{and the pairing vertex  (Eq. \eqref{eq:gapEqOrdered}) } contain 
contributions from the $q_0=0$ Matsubara mode, which diverge at the (QCP). In the following, we show that this contribution can be safely removed even in the ordered state, due to the Ising nature of the interaction. The procedure turns out to be analogous to the one in the disordered state, but for completeness we do the derivation step-by-step.

Within the framework of Eliashberg theory, the fermionic full Green’s function $\hat{G}_{l}$  and self-energy $\hat{\Sigma}_{l}$ in Nambu space can be expressed as \cite{MoonChubukov2010QuantumCritical}
\begin{equation}
 \begin{aligned}
        &\hat{\Sigma}_{l}(k_0)=-i\Sigma_{l}(k_0) \hat{\tau}_0-\Phi_{l}(k_0)\hat{\tau}_1 \, ,\\
        &\hat{G}_{l}^{-1}(\epsilon_k,k_0)=\hat{G}_{0}^{-1}(\epsilon_k,k_0)-\hat{\Sigma}_{l}(k_0)=i\tilde\Sigma_{l}(k_0)\hat{\tau}_0-\epsilon_k\hat{\tau}_3+\Phi_{l}(k_0)\hat{\tau}_1 \, .
         \label{eq:green_nambu}
 \end{aligned}
\end{equation}
Here, $\hat{\tau}$ are the Pauli matrices in the particle-hole space. $\epsilon_k$ represents the fermionic dispersion in the normal state. $\hat{G}_0$ is the Green's function for the non-interacting fermions. $\Phi_{l}(k_0)$ is the pairing vertex and $\Sigma_{l}$ is the regular self-energy (not in Nambu space). $\hat{\Sigma}_{l}$ can be written using $\hat{G}_{l}$  as

\begin{equation}
 \begin{aligned}
	\hat{\Sigma}_{l}(k_0)&=\frac{\bar{\lambda}}{ k_a^2} \frac{1}{(2\pi)^2}  T \sum_{q_0}  \int q D(\bar{q}) \left[\int d\theta \, \hat{G}_{l}(k_0,\epsilon_k)  \right]dq\,  \\
	-i\Sigma_{l}(k_0) \hat{\tau}_0-\Phi_{l}(k_0)\hat{\tau}_1  &=   \frac{\bar{\lambda}}{ k_a^2} \frac{1}{(2\pi)^2}  T \sum_{q_0}  \int q D(\bar{q}) \left[\int d\theta    \frac{-i\tilde{\Sigma}_{l}(k_0+q_0)\hat{\tau}_0-\epsilon_{\bm{k}+\bm{q}}\hat{\tau}_3+\Phi_{l}(k_0+q_0)\hat{\tau}_1 }{\vert \tilde{\Sigma}_{l}(k_0+q_0)\vert^2+\epsilon_{\bm{k}+\bm{q}}^2+\Phi_{l}^2(k_0+q_0)} \right]dq\,. \\
	 \label{eq:self_nambu}
 \end{aligned}
\end{equation}

Comparing the coefficients of the Pauli matrices in Eq.~\eqref{eq:self_nambu}, we obtain the following set of Eliashberg equations \cite{MoonChubukov2010QuantumCritical, ChubukovAbanovWangWu2020InterplaySuperconductivity}.
\begin{equation}
 \begin{aligned}
\Sigma_{l}(k_0)    &=\frac{\bar{\lambda}}{ k_a^2} \frac{1}{(2\pi)^2}  T \sum_{q_0}  \int q D(\bar{q}) \tilde{\Sigma}_{l}(k_0+q_0) \left[\int    \frac{ d\theta }{\vert \tilde{\Sigma}_{l}(k_0+q_0)\vert^2+\epsilon_{\bm{k}+\bm{q}}^2+\Phi_{l}^2(k_0+q_0)} \right]dq \\
&=\frac{\bar{\lambda}}{ k_a^2} \frac{1}{(2\pi)^2}  T \sum_{q_0}  \int q D(\bar{q}) \tilde{\Sigma}_{l}(k_0+q_0)   \frac{1}{(v_{Fl} \,q)^2} \left[\int_0^{2\pi} \frac{d\theta}{ \frac{\vert \tilde{\Sigma}_{l}(k_0+q_0) \vert^2+\Phi_{l}^2(k_0+q_0)} {(v_{Fl} \,q)^2}+ \cos^2\theta}   \right] dq \\
&=\frac{\bar{\lambda}}{ k_a^2} \frac{1}{2\pi}  T \sum_{q_0}  \int q D(\bar{q}) \tilde{\Sigma}_{l}(k_0+q_0)   \frac{1}{(v_{Fl} \,q)^2} \left[    \frac{1}{  \frac{\sqrt{\vert \tilde{\Sigma}_{l}(k_0+q_0) \vert^2+\Phi_{l}^2(k_0+q_0)}} {v_{Fl} \,q } \sqrt{1+\frac{\vert \tilde{\Sigma}_{l}(k_0+q_0) \vert^2+\Phi_{l}^2(k_0+q_0)} {(v_{Fl} \,q)^2}}}   \right] dq \\
&=\frac{\bar{\lambda}\nu_{F}}{ k_{Fl} a}  T \sum_{q_0}  \frac{d(q_0) \tilde{\Sigma}_{l}(k_0+q_0)}{\sqrt{\vert \tilde{\Sigma}_{l}(k_0+q_0) \vert^2+\Phi_{l}^2(k_0+q_0)}} \, .
	 \label{eq:Eliashberg_self}
 \end{aligned}
\end{equation}
Eq.~\eqref{eq:Eliashberg_self} are obtained using $\tilde{\Sigma}_{l}(k_0+q_0) \vert \ll  v_{Fl} \vert q \vert$. Similarly we obtain
\begin{equation}
 \begin{aligned}
\Phi_{l}(k_0)&=\frac{\bar{\lambda}\nu_{F}}{ k_{Fl} a}  T \sum_{q_0}  \frac{d(q_0) \Phi_{l}(k_0+q_0)}{\sqrt{\vert \tilde{\Sigma}_{l}(k_0+q_0) \vert^2+\Phi_{l}^2(k_0+q_0)}} \, .
	 \label{eq:Eliashberg_vertex}
 \end{aligned}
\end{equation}



At the QCP, the interaction kernel $d(q_0)$ acquires a diverging contribution at $q_0=0$
(see Eq.~\eqref{eq:dq0_integral}), corresponding to the thermal fluctuation
\cite{KleinKoziiRuhmanFernandes2023QuantumFerroelectricMetals, ChubukovSchmalian2005MasslessBoson}.
By analogy with nonmagnetic impurities, this static contribution does not suppress $T_c$
(Anderson’s theorem) \cite{Anderson1959DirtySuperconductors}. We therefore remove the thermal contribution
from the self-energy and pairing equations by subtracting the $q_0=0$ term in
Eqs.~\eqref{eq:Eliashberg_self} and \eqref{eq:Eliashberg_vertex}, following
Refs.~\cite{MillisSachdevVarma1988Inelastic, AbanovChubukovNorman2008GapAnisotropy, MoonChubukov2010QuantumCritical, ChubukovAbanovWangWu2020InterplaySuperconductivity, Saha2025Strong}.
\AK{
To this end, we introduce the gap function
\begin{equation}
\label{eq:delta-def-app}
    \Delta_l(k_0)=k_0 \frac{\Phi_{l}(k_0)}{\tilde{\Sigma}_{l}(k_0)}\,,
\end{equation}
which is invariant under the transformation
\begin{align}
\tilde{\Sigma}_{l}(k_0) &\rightarrow \tilde{\Sigma}_{l}(k_0)\,\bigl[1 - P_{l}(k_0)\bigr], \\
\Phi_{l}(k_0) &\rightarrow \Phi_{l}(k_0)\,\bigl[1 - P_{l}(k_0)\bigr],
\label{eq:thermal}
\end{align}
with
\begin{equation}
P_{l}(k_0) =
\frac{\bar{\lambda}\,\nu_{F}}{k_{Fl} a}\; T \,
\frac{d(0)}{\sqrt{\tilde{\Sigma}_{l}^2(k_0) + \Phi_{l}^2(k_0)}} \,,
\end{equation}
which removes the divergent $d(0)$ part. Recasting $\Phi_l$ in terms of the new gap function we obtain,
\begin{flalign}
 	\Phi_{l}(k_0) = \frac{\tilde{\Sigma}_{l}(k_0)\Delta_l(k_0)}{k_0}=\Delta_l(k_0)+\frac{\bar{\lambda}\nu_{F}}{ k_{Fl} a}  T \sum_{q_0}  d(q_0)\frac{\Delta_l(k_0) \frac{k_0+q_0}{k_0}}{\sqrt{(k_0+q_0)^2+\Delta^2_l(k_0+q_0)}}\nonumber\\
    = \frac{\bar{\lambda}\nu_{F}}{ k_{Fl} \,a}  T \sum_{q_0} d(q_0) \frac{\Delta_l(k_0+q_0)}{\sqrt{(k_0+q_0)^2+\Delta_l^2(k_0+q_0)}}\,.
	 \label{eq:Eliashberg_self_del}
 \end{flalign}
Here, the first line is obtained from Eq. \eqref{eq:delta-def-app}, and the second from Eq. \eqref{eq:Eliashberg_vertex}. Clearly, the $q_0=0$ term drops out, and we are left with
}
\begin{equation}
 \begin{aligned}	\Delta_l(k_0)&=\frac{\bar{\lambda}\nu_{F}}{ k_{Fl} \,a}  T \sum_{q_0 \ne 0} d(q_0) \left(  \frac{\Delta_l(k_0+q_0)}{\vert k_0+q_0 \vert} -   \text{sgn}(k_0+q_0) \frac{\Delta_l(k_0)}{k_0} \right)\,.
	 \label{eq:Eliashberg_vertex_thermal}
 \end{aligned}
\end{equation}
After substituting $k_0+q_0=p_0$, Eq.~\eqref{eq:Eliashberg_vertex_thermal} leads to the following eigenvalue equation
\begin{equation}
\Delta_l(k_0)=
\frac{\displaystyle 
\frac{\bar{\lambda}\,\nu_{F}}{k_{Fl}a}\;
T \sum_{p_0 \ne k_0} 
\frac{1}{p_0}\bigl[d(p_0-k_0)+d(p_0+k_0)\bigr]\,
\Delta_l(p_0)}
{\displaystyle 
1+\frac{\bar{\lambda}\,\nu_{F}}{k_{Fl}a}\;
T \,\frac{1}{k_0}
\sum_{p_0^\prime \ne k_0}
\bigl[d(p_0^\prime -k_0)-d(p_0^\prime +k_0)\bigr]}\,,
\label{eq:eliashberg_gapeq_matrix1}
\end{equation}
\AK{This is precisely the same form as the equation for the disordered case, but with layer-dependent electronic parameters. One may verify that the argument goes through for a layer dependent DOS as well.}

\section{\label{app:delta} 
$T_c$ splitting in the ordered state}
In this appendix, we present the analytical steps leading to the expression for the layer splitting $\Delta T_c/T_c$,
%
%
%
%
starting from the linearized pairing equation on layer $l$, Eq.~\eqref{eq:paringdordered},
\begin{equation}
\Phi_{l}(k_0)
=\frac{\bar\lambda\, \nu_{F}}{k_{Fl}a}\;
T\sum_{q_{0,n}}
\frac{\Phi_{l}(k_0+q_{0,n})\,d(q_{0,n})}{\bigl|\tilde\Sigma_{l}(k_0+q_{0,n})\bigr|}\, .
\label{eq:paringdordered_appendix}
\end{equation}
Here (only in this appendix) we explicitly write the bosonic Matsubara frequency as $q_{0,n} = 2\pi n T$, with integer index $n$, to emphasize that 
we sum over $n$ to determine $T_c$. 
\AK{In what follows, to make the calculations more transparent, we will just neglect the $n=0$ term rather than removing it as we discussed above. In addition, we approximate the pairing function $\Phi_l$ to be constant, neglecting its large frequency decay. It can be verified that this does not change the qualitative result. Explicitly, then, we have}
\begin{equation}
1 \;=\; \frac{\bar\lambda\, \nu_{F}}{k_{Fl}a}\,T\sum_{n\ge 1} d_r(q_{0,n})\,f_l(q_{0,n}) \,,
\label{eq:gap-criterion}
\end{equation}
where $f_l = |\tilde\Sigma_l(q_{0,n})|^{-1}$ accounts for the fermionic contribution to the pairing kernel, and $d_r$ accounts for the bosonic part.

\subsection{Dimensionless form of the gap condition}
To estimate the layer splitting, it is convenient to cast Eq.~\eqref{eq:gap-criterion} into a fully dimensionless form. Following Appendix~\ref{app:ordered} in the ordered state, we can write
\begin{equation}
d_r(q_0) \;=\; \int_0^{\infty}\!\!\frac{dq}
{\,2|r| + q^2 + \Pi (q_0)}\,.
\label{eq:dr-def}
\end{equation}

Expanding the bosonic propagator near the QCP and keeping terms to linear order in small $r$, we obtain
\begin{equation}
\frac{1}{2r + q^2 + \Pi}
\;=\;
\frac{1}{q^2 + \Pi}
\;-\;
\frac{2r}{(q^2 + \Pi)^2}
\;+\; O(r^2)\,.
\label{eq:D-expand}
\end{equation}
After inserting Eq.~\eqref{eq:D-expand} into Eq.~\eqref{eq:dr-def} and performing the momentum integral, we separate the pairing kernel into a critical part $d_0$ ($r=0$) and a mass correction $\Delta d$ (finite $r$),
\begin{equation}
d_r(q_0) = d_0(q_0) + \Delta d(q_0)\,,
\label{eq:dr}
\end{equation}
with
\begin{equation}
d_0(q_{0,n}) = \frac{2\pi}{3\sqrt{3}}\,
\frac{1}{\gamma_{eff}^{1/3}(2\pi n T)^{1/3}}\,,
\qquad
\Delta d(q_{0,n}) = -\frac{2 r}{3\gamma_{eff}}\,\frac{1}{2\pi n T}\,,
\end{equation}
where Landau damping term $\gamma_{eff}= 4\bar \lambda \nu_F/k_av_{eff}$. The fermionic contribution can also be re-written as
\begin{equation}
f_l(q_{0,n})
= \frac{1}{|\tilde\Sigma_l(q_{0,n})|}
= \frac{1}{\omega_{0,l}^{1/3}(2\pi n T)^{2/3}}
\frac{1}{1+a_l n^{1/3}}\,,
\label{eq:fl}
\end{equation}
where $\omega_{0,l}$ can be obtained from Eq. \eqref{eq:k23} by replacing $k_F\to k_{F,l}$, etc., and $a_l$ is a dimensionless temperature variable to solve Eq. \eqref{eq:gap-criterion}:
\begin{equation}
a_l \equiv \left(\frac{2\pi T}{\omega_{0,l}}\right)^{1/3}\,.
\label{eq:al}
\end{equation}
Since $T_{c,l}\sim \omega_{0,l}$, $a_l$ is of order unity.
Using Eqs.~\eqref{eq:dr} and \eqref{eq:fl} in Eq.~\eqref{eq:gap-criterion}, each factor of $T$
cancels against the Matsubara spacing $2\pi nT$, and the gap equation becomes a purely dimensionless equation,
\begin{equation}
1
= \frac{\bar\lambda\, \nu_{F}}{k_{Fl}a}
\left[
\frac{1}{3\sqrt{3}}\frac{1}{\gamma_{eff}^{1/3}\omega_{0,l}^{1/3}}\,S_1(a_l)
-\frac{1}{3\pi}\frac{r}{\gamma_{eff}\omega_{0,l}}\,a_l^{-2}\,S_{5/3}(a_l)
\right]\,,
\label{eq:dimless-eq}
\end{equation}
with
\begin{equation}
S_\nu(a) = \sum_{n\ge 1} \frac{n^{-\nu}}{1+a n^{1/3}}\,.
\end{equation}
All microscopic information is contained in the terms
$\gamma_{eff},\, \omega_{0,l}$,$(\bar\lambda\, \nu_{F}/k_{Fl}\, a)$, and $r$, while $S_1$ and $S_{5/3}$ are
numerical functions of $a_l$.

\subsection{Linear shift of $T_c$ at small $r$}

At the FE quantum critical point ($r=0$) the second term in Eq.~\eqref{eq:dimless-eq}
vanishes, and one obtains
\begin{equation}
1
= \frac{\bar\lambda\, \nu_{F}}{k_{Fl}a}\, \frac{1}{3\sqrt{3}}
\frac{1}{\gamma_{eff}^{1/3}\omega_{0,l}^{1/3}}\,
S_1(a_{0,l})\,,
\label{eq:a0-def}
\end{equation}
which determines the reference dimensionless temperature $a_{0,l}$. Since $\gamma_{eff}^{1/3}\,\omega_{0,l}^{1/3}\sim \bar{\lambda}\nu_F/(k_F a)$, the equation is completely dimensionless, confirming that $T_c \sim \omega_0$ and that $a_l = O(1)$.

For small but nonzero $r$ we can expand $a_l$ (because this is related to $\delta T_l$) in Eq. \eqref{eq:al} 
\begin{equation}
a_l = a_{0,l} + \delta a_l\,,
\end{equation}
and define
\begin{equation}
F_l(a_l,r)
\equiv
\frac{\bar\lambda\, \nu_{F}}{k_{Fl}a}\left[
\frac{1}{3\sqrt{3}}\frac{1}{\gamma_{eff}^{1/3}\omega_{0,l}^{1/3}}\,S_1(a_l)
-\frac{1}{3\pi}\frac{r}{\gamma_{eff}\omega_{0,l}}\,a_l^{-2}\,S_{5/3}(a_l)
\right]-1\,.
\end{equation}
By construction $F_l(a_l,r)=0$ at the transition, and Eq.~\eqref{eq:a0-def}
implies $F_l(a_{0,l},0)=0$.
Linearizing around $(a_{0,l},0)$, we obtain
\begin{equation}
0 = F_l(a_{0,l},0)
+ \left.\frac{\partial F_l}{\partial a}\right|_0 \delta a_l
+ \left.\frac{\partial F_l}{\partial r}\right|_0 r
+ O(\delta a_l^2,r^2,r\delta a_l)\,,
\end{equation}
so that
\begin{equation}
\delta a_l
= -\left.
\frac{\partial F_l/\partial r}{\partial F_l/\partial a}
\right|_{(a_{0,l},0)} r\,.
\end{equation}
A short calculation using Eq.~\eqref{eq:dimless-eq} gives
\begin{equation}
\left.\frac{\partial F_l}{\partial r}\right|_0
= -\frac{\bar\lambda\, \nu_{F}}{k_{Fl}a}\,\frac{1}{3\pi}\frac{1}{\gamma_{eff}\omega_{0,l}}
\,a_{0,l}^{-2}\,S_{5/3}(a_{0,l})\,,
\end{equation}
\begin{equation}
\frac{\partial F_l}{\partial a}
= \frac{\bar\lambda\, \nu_{F}}{k_{Fl}a} \frac{1}{3\sqrt{3}}\frac{1}{\gamma_{eff}^{1/3}\omega_{0,l}^{1/3}}\,S_1'(a_l)\,,
\qquad
S_1'(a) = -\sum_{n\ge 1}\frac{n^{-2/3}}{(1+a n^{1/3})^2}\,,
\end{equation}
so that
\begin{equation}
\delta a_l
= \frac{\sqrt{3}}{\pi}\,
\frac{r}{\gamma_{eff}^{2/3}\omega_{0,l}^{2/3}}\,
\frac{a_{0,l}^{-2}\,S_{5/3}(a_{0,l})}{S_1'(a_{0,l})}\,.
\end{equation}
Since $T=(\omega_{0,l}/2\pi)a_l^3$ from Eq. \eqref{eq:al}, the relative shift of the critical temperature
is
\begin{equation}
\frac{\delta T_l}{T_0}
= 3\,\frac{\delta a_l}{a_{0,l}}
=
\frac{3\sqrt{3}}{\pi}\,
\frac{r}{\gamma_{eff}^{2/3}\omega_{0,l}^{2/3}}\,
\frac{a_{0,l}^{-3}\,S_{5/3}(a_{0,l})}{S_1'(a_{0,l})}\,,
\label{eq:deltaT-layer}
\end{equation}
which is linear in $r$ for each individual layer.

\subsection{Layer splitting and $r^2$ scaling}

The layer splitting of the transition temperature is
\begin{equation}
\Delta T_c \equiv T_c^t - T_c^b\,,
\qquad
T_c^l = T_0 + \delta T_l\,.
\end{equation}
For small shifts we may write
\begin{equation}
\frac{\Delta T_c}{T_c}
\simeq \frac{\delta T_t - \delta T_b}{T_0}\,.
\end{equation}
Using Eq.~\eqref{eq:deltaT-layer}, this becomes
\begin{equation}
\frac{\Delta T_c}{T_c}
=
\frac{3\sqrt{3}}{\pi}\,r\,
\left[
\frac{1}{\gamma_{eff}^{2/3}\omega_{0,t}^{2/3}}
\frac{a_{0,t}^{-3}\,S_{5/3}(a_{0,t})}{S_1'(a_{0,t})}
-
\frac{1}{\gamma_{eff}^{2/3}\omega_{0,b}^{2/3}}
\frac{a_{0,b}^{-3}\,S_{5/3}(a_{0,b})}{S_1'(a_{0,b})}
\right]\,.
\label{eq:DeltaTc-exact}
\end{equation}

Since the two layers are exactly equivalent at the QCP ($r=0$), all parameters entering Eq.~\eqref{eq:DeltaTc-exact} 
coincide.
For $r<0$, the FE order weakly breaks this symmetry and generates a small layer asymmetry, and since all the relevant parameters are analytic functions of $r$, the asymmetry
$\propto r$ at small $r$. The dimensionless expression inside the square brackets in Eq.~\eqref{eq:DeltaTc-exact} is a smooth function of these parameters, so it also vanishes at $r=0$ and its leading nonzero contribution is linear in $r$. Combined with the explicit prefactor of $r$ in Eq.~\eqref{eq:DeltaTc-exact}, this implies
\begin{equation}
\frac{\Delta T_c}{T_c} \approx \mathcal{C} r^2+ O(r^3)\,,
\end{equation}
up to higher-order corrections, this is exactly Eq. \eqref{eq:deltaTc/Tc}. Here $\mathcal{C}$ is a dimensionless constant that depends only on the QCP
solution $a_0$ and on the layer parameters for $r=0$.
This shows that the layer splitting of $T_c$ is parametrically small and scales as $O(r^2)$ near the FE quantum critical point see Fig. \ref{Fig:Tc_coupling}b.


\section{\label{app:MoTe2}Estimate of $T_c$ for bilayer MoTe$_2$}

\begin{figure}[H]
    \centering
\includegraphics[width=.5\linewidth]{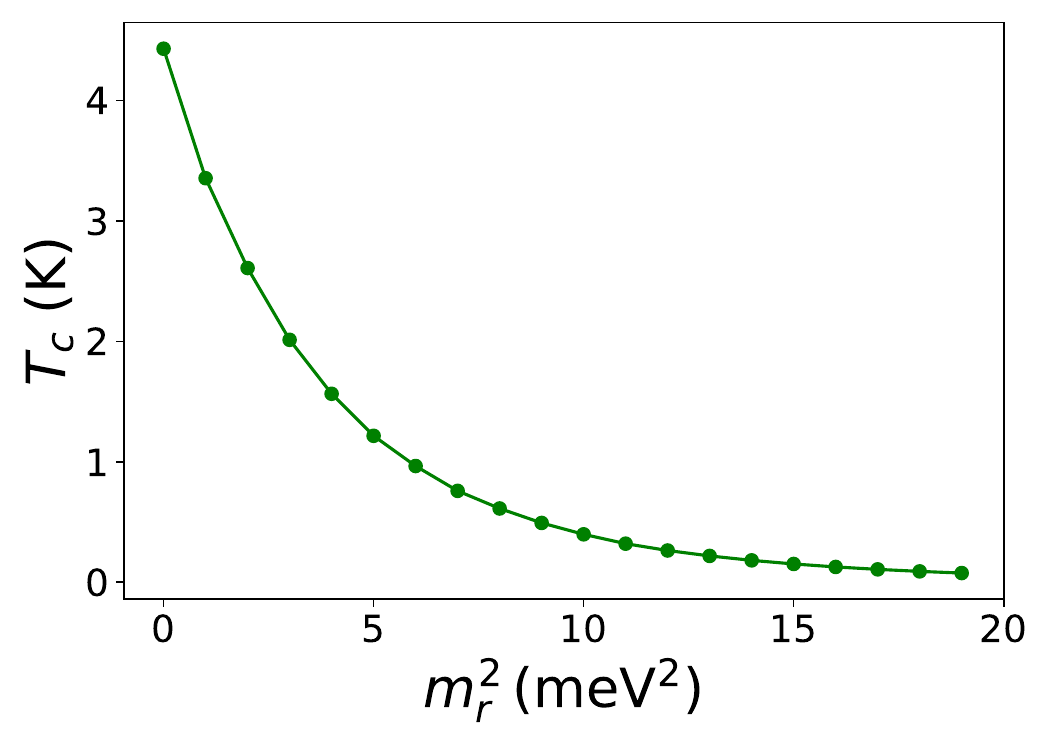}
   \caption{$T_c$ as a function of the renormalized phonon mass $m_r^2$ in bilayer MoTe$_2$ within the disordered phase. For $\bar{\lambda} = 225~\mathrm{meV}$ the calculation yields $T_c \sim 4.5~\mathrm{K}$, consistent with experiment.}
    \label{Fig:mote2_disorder}
\end{figure}

In this appendix we estimate the superconducting transition temperature of bilayer MoTe$_2$ using the Eliashberg gap equation, Eq.~\eqref{eq:eliashberg_gapeq_matrix}, with the material parameters listed in Table~\ref{tab:mote2-params}. Restricting to the disordered (non–ferroelectric) phase, we find that an effective electron–phonon coupling strength $\bar{\lambda} = 225~\mathrm{meV}$ yields a transition temperature $T_c \sim 4.5~\mathrm{K}$, on order of the experiment \cite{Jindal2023}. The dependence of $T_c$ on the renormalized phonon mass $m_r^2$ is shown in Fig.~\ref{Fig:mote2_disorder}, and illustrates the rapid enhancement of pairing as the sliding mode softens. A corresponding analysis in the ordered sliding–FE phase is deferred to future work, as several key microscopic parameters for this regime are not currently available for MoTe$_2$.

\section{\label{app:numerics}Numerical convergence}

Here we briefly demonstrate the  convergence of our numerical solutions. The two numerically nontrivial cutoffs in our algorithm are the upper momentum cutoff in the computation of $d(q_0)$, Eq. \eqref{eq:dq0_integral},  and the maximum Matsubara frequency in e.g. \eqref{eq:pairing-matsubara}. 
\begin{figure}[H]
    \centering
    \includegraphics[width=1\linewidth]{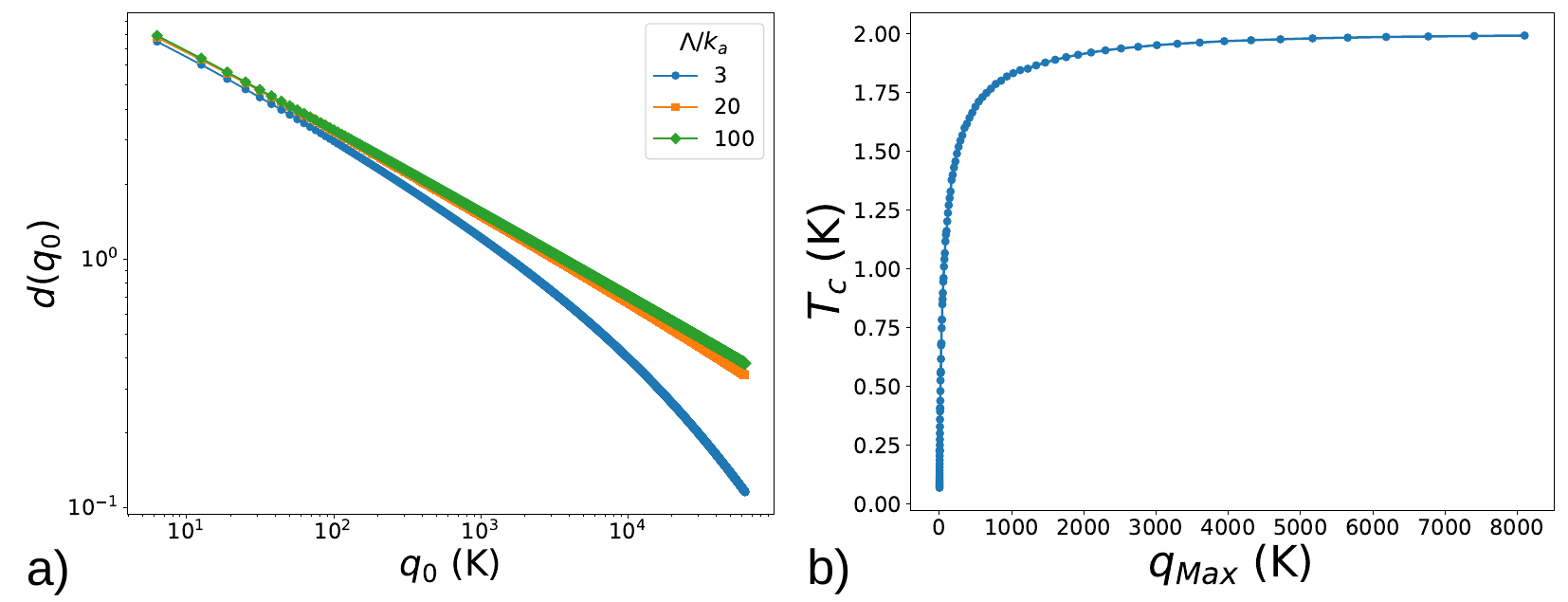}
    \caption{a) $d(q_0)$ vs $q_0$ for different cutoff momenta $\Lambda/k_a$. b) $T_c$ vs $q_{Max}$.}
    \label{Fig:numerics}
\end{figure}
We present the frequency-dependent integration $d(q_0)$ [Eq.~\eqref{eq:dq0_integral}] as a function of Matsubara frequency $q_0$ (in units of K), for different cutoff values in Fig. \ref{Fig:numerics}. Rapid convergence is clear in the figure. 
The Matsubara frequency cutoff in Eq.~\eqref{eq:eliashberg_gapeq_matrix1} is set by $E_F/k_B \approx 7000~\text{K}$, which is far higher than is necessary for numerical convergence, as shown in Fig. \ref{Fig:numerics}. 
Therefore, the superconducting transition temperatures $T_c$ presented in the paper do not depend on the specific choice of cutoff (momentum, frequency).

\section{Calculation of Josephson coupling}
\label{sec:app-jos}
Let the momenta in the top and bottom layers be represented by $k$ and $q$, respectively. 
We write the generalized hopping hamiltonian in a slightly different form
\begin{equation}
H_{tb} 
=
\sum_{kq\al\bt} M_{tk\al}^{bq\bt} \ \psi_{t,k,\al}^{\dg}  \psi_{b,q,\bt} +
\left(M_{tk\al}^{bq\bt}\right)^* \ \psi_{b,q,\bt}^{\dg}  \psi_{t,k,\al}^{}\,,
\end{equation}
where the hopping $M$ is a complex matrix, and can be  decomposed in $\tau$ and $\s$ basis, as $M_{l\al}^{l'\bt}=\sum_{ij}\Lambda_{ij}\tau_i^{ll'} \s_j^{\al\bt}$, with $i=t,b$ and $j=0,x,y,z$ and ignoring the momentum dependency. We keep the momentum indices explicit for the discussion below. 
The time-reversal operator acts as follows
\[ \mathcal T^{-1} \sum_{kq\al\bt} M_{tk\al}^{bq\bt} \ \psi_{t,k,\al}^{\dg}  \psi_{b,q,\bt} \mathcal T = \sum_{kq\al\bt} \mathrm{sign}(\al \bt) (M_{tk\al}^{bq\bt})^* \ \psi_{t,-k,-\al}^{\dg}  \psi_{b,-q,-\bt} \,, \]
such that TRS implies that 
\[M_{tk\al}^{bq\bt}=\mathrm{sign}(\al \bt) (M_{t,-k,-\al}^{b,-q,-\bt})^* \,.\]
On the other hand, the inversion operator gives
\[ \mathcal I^{-1} \sum_{kq\al\bt} M_{tk\al}^{bq\bt} \ \psi_{t,k,\al}^{\dg}  \psi_{b,q,\bt} \mathcal I = \sum_{kq\al\bt}M_{tk\al}^{bq\bt} \ \psi_{b,-k,\al}^{\dg}  \psi_{t,-q,\bt}\,,
\]
and as a result the inversion symmetry implies 
\[M_{tk\al}^{bq\bt}=M_{b,-k,\al}^{t,-q,\bt}\,. \]



\begin{figure}[H]
    \centering
    \includegraphics[width=1\linewidth]{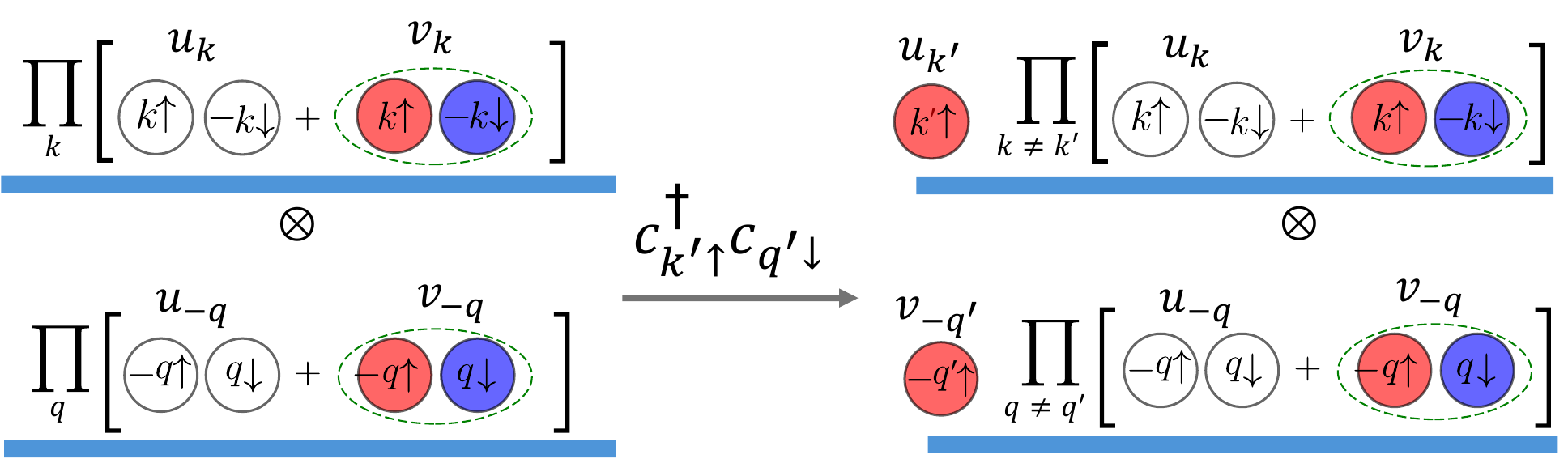}
    \caption{The combined BCS wave function of the two layers is shown on the left, with $k$ and $q$ denoting the momenta in the top and bottom layers, respectively. Each bracketed term represents a superposition of an unoccupied $k\ur,-k\dr$ state with amplitude $u_k$ (depicted by empty circles) and an occupied Cooper-pair state with amplitude $v_k$ (represented by the enclosed green curve).  The hopping $c^\dg_{k^\prime \ur}c_{q^\prime \dr}$ creates one unpaired spin-$\ur$ particle in each layer. The same unpaired particles can equivalently be generated by the operator $c^\dg_{-q^{\prime}\ur}c_{-k^{\prime}\dr}$. Because, to create an upaired particle in the $k^{\prime}\ur$ states, one can either occupy the empty state (e.g., by the former hopping term) associated with $u_{k^{\prime}}$, or annihilate the $-k^{\prime}\dr$ state (e.g., by the latter hopping term) from the Cooper-pair associated with $v_{k^{\prime}}$, and similarly for the unpaired particles in the $-q^{\prime}\ur$ state.}
    \label{Fig:Josephson_hopping}
\end{figure}

For singlet pairing, the BCS wave function in the top layer is $\Psi_t=\prod_k(u_{tk}+v_{tk} \psi_{t,k\ur}^\dg \psi_{t,-k\dr}^\dg)\ket{0}$, with an analogous form $\Psi_b$ for the bottom layer. Acting with $\psi_{tk\al}^\dagger \psi_{bq\bt}$ on the product state $\Psi = \Psi_t \otimes \Psi_b$ creates one quasiparticle in each layer with momenta and spins $(k,\alpha)$ and $(-q,\bar{\beta})$, where $\bar\beta$ denotes the spin opposite to $\beta$. 
For example,
$
\psi^\dg_{t,k\ur} \psi_{b,q\dr} \Psi= - u_{tk}v_{bq} \psi^{\dg}_{t,k\ur} \psi_{b,-q\ur}^\dg (\text{rest of }\Psi)
$, 
as illustrated in Fig.~\ref{Fig:Josephson_hopping}. 
 

Among all hopping terms, only the operator $\psi_{b,-q\bar{\bt}}^{\dg}  \psi_{t,-k\bar{\al}}$ creates the same excited state, leading to a total amplitude $\big[\mathrm{sgn}(\al\bt)M_{tk\al}^{bq\bt} \ u_{tk} v_{bq} +
M_{b,-q\bar\bt}^{t,-k\bar\al} \ u_{bq} v_{tk} 
\big]$.
The corresponding second-order perturbative correction to the ground state energy is:  
\begin{align}
    \delta E_2 \propto -\sum |\mathrm{sgn}(\al\bt) M_{tk\al}^{bq\bt} \ u_{tk} v_{bq} +
M_{b-q\bar\bt}^{t-k\bar\al} \ u_{bq} v_{tk}|^2\,, 
\end{align}
where the proportionality constant is positive and the gap is given by $\Delta_{t(b)}\propto u_{kt(bq)} v_{tk(bq)}^*$. Consequently, the phase-sensitive term in the square,
\[
\delta E_2\sim
-\mathrm{sgn}(\al\bt) [M_{tk\al}^{bq\bt} (M_{b,-q\bar\bt}^{t,-k\bar\al})^* \Delta_t \Delta_b^* + c.c]\,, 
\]
determines the relative phase locking between $\Delta_t$ and $\Delta_b$. Suppose, 
\[
M_{tk\ur}^{bq\dr} (M^{b,-q\dr}_{t,-k\ur})^*= e^{-i\phi_h}|M|^2\,,  \quad
\Delta_t \Delta_b^*=e^{i\phi}|\Delta|^2\,,
\]
then
\begin{align}
    \delta E_2 \propto \mathrm{cos}(\phi-\phi_h) |\Lambda\Delta|^2\,,
\end{align}
which is minimized when $\phi=\phi_h+\pi$. 

For spin preserving flipping, if 
\[
M_{tk\ur}^{bq\ur} (M^{b,-q\dr}_{t,-k\dr})^*= e^{-i\phi_h}|M|^2,  \quad
\Delta_t \Delta_b^*=e^{i\phi}|\Delta|^2\,,
\]
then
\begin{align}
    \delta E_2 \propto -\mathrm{cos}(\phi-\phi_h) |\Lambda\Delta|^2\,,
\end{align}
which is minimized when $\phi=\phi_h$. Details about $\phi$ for various possibilities under $\mt{I}$ and $\mt{T}$ operators are shown in Table~\ref{Tab:hop_withoutMz} and Table~\ref{Tab:phi_withoutMz}.

The case with mirror-z symmetry ($\mt{M}_z$) is shown in Table~\ref{Tab:hop_with Mz} and Table~\ref{Tab:phi_with Mz}. The mirror symmetry does not change the conclusions obtained with $\mt{I}$ and $\mt{T}$.
\begin{table}[H]
\centering
\begin{tabular}{|c|c|c|}
\hline
Hopping matrices  & $\mt{I}$ & $\mt{T}$ \\
\hline \hline
$\tau_x\sigma_x$, $\tau_x\sigma_y$, $\tau_x\sigma_z$  & + & - \\
\hline
$\tau_x\sigma_0$  & + & + \\
\hline
$\tau_y\sigma_x$, $\tau_y\sigma_y$, $\tau_y\sigma_z$  & - & + \\
\hline
$\tau_y\sigma_0$  & - & - \\
\hline
\end{tabular}
\caption{The table with inversion and time-reversal from the main paper are reproduced here for the later part.}
\label{Tab:hop_withoutMz}
\end{table}

\begin{table}[ht]
\begin{tabular}{|c|c||cccc||ccc|}
\hline
\multirow{2}{*}{$\mt{I}$} & \multirow{2}{*}{$\mt{T}$} & \multicolumn{4}{c||}{Amplitude for spin flip}  & \multicolumn{3}{c|}{Amplitude for no flip}      \\ 
\cline{3-9}  &  & \multicolumn{1}{c|}{$t\ur\to b\dr$} & \multicolumn{1}{c|}{$b\ur\to t\dr$} & \multicolumn{1}{c|}{$\phi_h$} & $\phi$ & \multicolumn{1}{c|}{$t\ur\to b\ur$} & \multicolumn{1}{c|}{$b\dr\to t\dr$} & $\phi_h=\phi$ \\ 
\hline  \hline
 + & -  & \multicolumn{1}{c|}{$\Lambda_{xx}-i\Lambda_{xy}$} & \multicolumn{1}{c|}{$\Lambda_{xx}-i\Lambda_{xy}$} & \multicolumn{1}{c|}{0} & $\pi$  & \multicolumn{1}{c|}{$\Lambda_{xz}+\Lambda_{x0}$} & \multicolumn{1}{c|}{$-\Lambda_{xz}+\Lambda_{x0}$}  & \multicolumn{1}{c|}{\makecell{$0$ if $\Lambda_{xz}<\Lambda_{x0}$ \\ $\pi$ if $\Lambda_{xz}>\Lambda_{x0}$}}     \\
\hline
+ & + & \multicolumn{1}{c|}{} & \multicolumn{1}{c|}{} & \multicolumn{1}{c|}{} &  & \multicolumn{1}{c|}{$\Lambda_{x0}$} &
\multicolumn{1}{c|}{$\Lambda_{x0}$} & 0 \\
\hline
- & + & \multicolumn{1}{c|}{$-i\Lambda_{yx}-\Lambda_{yy}$} & \multicolumn{1}{c|}{$i\Lambda_{yx}+\Lambda_{yy}$} & \multicolumn{1}{c|}{$\pi$} & 0 & \multicolumn{1}{c|}{$-i\Lambda_{yz}+\Lambda_{x0}$} &
\multicolumn{1}{c|}{$-i\Lambda_{yz}+\Lambda_{x0}$} & 0 \\
\hline
- & -  & \multicolumn{1}{c|}{\makecell{$\Lambda_{xx}-\Lambda_{yy}$ \\ $- i(\Lambda_{xy}+\Lambda_{yx})$}} &
\multicolumn{1}{c|}{\makecell{$\Lambda_{xx}+\Lambda_{yy}$ \\ $-i(\Lambda_{xy}-\Lambda_{yx})$}} &
\multicolumn{1}{c|}{$\phi_h$} & $\phi_h+\pi$ &
\multicolumn{1}{c|}{\makecell{$\Lambda_{xz}+\Lambda_{x0}$ \\ $-i(\Lambda_{yz}+\Lambda_{y0})$}}
&
\multicolumn{1}{c|}{\makecell{$-\Lambda_{xz}+\Lambda_{x0}$ \\ $-i(\Lambda_{yz}-\Lambda_{y0})$}} & $\phi_h$ \\
\hline
\end{tabular}
\caption{Table with Inversion and time-reversal symmetry. }
\label{Tab:phi_withoutMz}
\end{table}

\begin{table}[ht]
\centering
\begin{tabular}{|c|c|c|c|}
\hline
Hopping matrices  & $\mt{I}$ & $\mt{T}$ & $\mt{M}_z$\\
\hline \hline
$\tau_x\sigma_x$, $\tau_x\sigma_y$  & + & - & - \\
\hline
$\tau_x\sigma_z$  & + & - & + \\
\hline
$\tau_x\sigma_0$  & + & + & +\\
\hline
$\tau_y\sigma_x$, $\tau_y\sigma_y$  & - & + & +\\
\hline
$\tau_y\sigma_z$  & - & + & -\\
\hline
$\tau_y\sigma_0$  & - & - & -\\
\hline
\end{tabular}
\caption{“$+$” (“$-$”) indicates that the corresponding interlayer hopping matrix is even (odd) under inversion, time reversal or Mirror-z operation. }
\label{Tab:hop_with Mz}
\end{table} 
\begin{table}[H]
\begin{tabular}{|c|c|c||cccc||ccc|}
\hline
\multirow{2}{*}{$\mt{I}$} & \multirow{2}{*}{$\mt{T}$} & \multirow{2}{*}{$\mt{M}_z$} & \multicolumn{4}{c||}{Amplitude for spin flip}  & \multicolumn{3}{c|}{Amplitude for no flip}      \\ 
\cline{4-10}  &  &  & \multicolumn{1}{c|}{$t\ur\to b\dr$} & \multicolumn{1}{c|}{$b\ur\to t\dr$} & \multicolumn{1}{c|}{$\phi_h$} & $\phi$ & \multicolumn{1}{c|}{$t\ur\to b\ur$} & \multicolumn{1}{c|}{$b\dr\to t\dr$} & $\phi_h=\phi$ \\ 
\hline  \hline
 + & -   & -  & \multicolumn{1}{c|}{$\Lambda_{xx}-i\Lambda_{xy}$} & \multicolumn{1}{c|}{$\Lambda_{xx}-i\Lambda_{xy}$} & \multicolumn{1}{c|}{0} & $\pi$  & \multicolumn{1}{c|}{} & \multicolumn{1}{c|}{}  & \multicolumn{1}{c|}{}     \\
\hline
+ & -  & + & \multicolumn{1}{c|}{} & \multicolumn{1}{c|}{}       & \multicolumn{1}{c|}{}         &   & \multicolumn{1}{c|}{$\Lambda_{xz}+\Lambda_{x0}$}       & \multicolumn{1}{c|}{$-\Lambda_{xz}+\Lambda_{x0}$}       & \multicolumn{1}{c|}{\makecell{$0$ if $\Lambda_{xz}<\Lambda_{x0}$ \\ $\pi$ if $\Lambda_{xz}>\Lambda_{x0}$}}     \\ 
\hline
+ & + & + & \multicolumn{1}{c|}{} & \multicolumn{1}{c|}{} & \multicolumn{1}{c|}{} &  & \multicolumn{1}{c|}{$\Lambda_{x0}$} &
\multicolumn{1}{c|}{$\Lambda_{x0}$} & 0 \\
\hline
- & + & + & \multicolumn{1}{c|}{$-i\Lambda_{yx}-\Lambda_{yy}$} & \multicolumn{1}{c|}{$i\Lambda_{yx}+\Lambda_{yy}$} & \multicolumn{1}{c|}{$\pi$} & \multicolumn{1}{c||}{0} & \multicolumn{1}{c|}{} &
\multicolumn{1}{c|}{} &  \\
\hline
- & + & -& \multicolumn{1}{c|}{} & \multicolumn{1}{c|}{} & \multicolumn{1}{c|}{} &  &
\multicolumn{1}{c|}{$-i\Lambda_{yz}+\Lambda_{x0}$} &
\multicolumn{1}{c|}{$-i\Lambda_{yz}+\Lambda_{x0}$} & 0 \\
\hline
- & - & - & \multicolumn{1}{c|}{\makecell{$\Lambda_{xx}-\Lambda_{yy}$ \\ $- i(\Lambda_{xy}+\Lambda_{yx})$}} &
\multicolumn{1}{c|}{\makecell{$\Lambda_{xx}+\Lambda_{yy}$ \\ $-i(\Lambda_{xy}-\Lambda_{yx})$}} &
\multicolumn{1}{c|}{$\phi_h$} & $\phi_h+\pi$ &
\multicolumn{1}{c|}{\makecell{$\Lambda_{xz}+\Lambda_{x0}$ \\ $-i(\Lambda_{yz}+\Lambda_{y0})$}}
&
\multicolumn{1}{c|}{\makecell{$-\Lambda_{xz}+\Lambda_{x0}$ \\ $-i(\Lambda_{yz}-\Lambda_{y0})$}} & $\phi_h$ \\
\hline
\end{tabular}
\caption{Table with inversion, time-reversal and mirror symmetry. }
\label{Tab:phi_with Mz}
\end{table}

\twocolumngrid
\bibliography{ref}

@article{almoalem2025mixed,
  title={Mixed Triplet-Singlet Order Parameter in Decoupled Superconducting 1H Monolayers of Transition-Metal Dichalcogenides},
  author={Almoalem, Avior and Kunhiparambath, Sajilesh and Gofman, Roni Anna and Nitzav, Yuval and Mangel, Ilay and Ragoler, Nitzan and Fujii, Jun and Vobornik, Ivana and Bertran, Francois and Kanigel, Amit and others},
  journal={arXiv preprint arXiv:2509.13303},
  year={2025}
}

@article{kumar2023first,
  title={First-order quantum phase transition in the hybrid metal--Mott insulator transition metal dichalcogenide {4Hb-TaS$_2$}},
  author={Kumar Nayak, Abhay and Steinbok, Aviram and Roet, Yotam and Koo, Jahyun and Feldman, Irena and Almoalem, Avior and Kanigel, Amit and Yan, Binghai and Rosch, Achim and Avraham, Nurit and others},
  journal={Proceedings of the National Academy of Sciences},
  volume={120},
  number={43},
  pages={e2304274120},
  year={2023},
  publisher={National Academy of Sciences}
}

@article{Silber2024,
  author    = {Silber, I. and Mathimalar, S. and Mangel, I. and Nayak, A. K. and
               Green, O. and Avraham, N. and Beidenkopf, H. and Feldman, I. and
               Kanigel, A. and Klein, A. and Goldstein, M. and Banerjee, A. and
               Sela, E. and Dagan, Y.},
  title     = {Two-component nematic superconductivity in {4Hb-TaS$_2$}},
  journal   = {Nature Communications},
  year      = {2024},
  volume    = {15},
  number    = {1},
  pages     = {824},
  doi       = {10.1038/s41467-024-45169-3},
  url       = {https://doi.org/10.1038/s41467-024-45169-3},
  issn      = {2041-1723}
}

@article{persky2022magnetic,
  title={Magnetic memory and spontaneous vortices in a van der Waals superconductor},
  author={Persky, Eylon and Bj{\o}rlig, Anders V and Feldman, Irena and Almoalem, Avior and Altman, Ehud and Berg, Erez and Kimchi, Itamar and Ruhman, Jonathan and Kanigel, Amit and Kalisky, Beena},
  journal={Nature},
  volume={607},
  number={7920},
  pages={692--696},
  year={2022},
  publisher={Nature Publishing Group UK London}
}

@article{Kanigel_npj2024,
title = {Charge transfer and spin-valley locking in {4Hb-TaS$_2$}},
  author = {Almoalem, Avior and Gofman, Roni and Nitzav, Yuval and Mangel, Ilay and Feldman, Irena and Koo, Jahyun and Mazzola, Federico and Fujii, Jun and Vobornik, Ivana and Sanchez-Barriga, J. and Clark, Oliver J. and Plumb, Nicholas Clark and Shi, Ming and Yan, Binghai and Kanigel, Amit},
  journal = {npj Quantum Materials},
  volume = {9},
  issue = {1},
  pages = {36},
  year = {2024},
  month = {Apr},
  doi = {10.1038/s41535-024-00646-2},
  url = {https://doi.org/10.1038/s41535-024-00646-2}
}

@article{Crippa_ncom2024,
title = {Heavy fermions vs doped Mott physics in heterogeneous Ta-dichalcogenide bilayers},
  author = {Crippa, Lorenzo and Bae, Hyeonhu and Wunderlich, Paul and Mazin, Igor I. and Yan, Binghai and Sangiovanni, Giorgio and Wehling, Tim and Valentí, Roser},
  journal = {Nature Communications},
  volume = {15},
  issue = {1},
  pages = {1357},
  year = {2024},
  month = {Feb},
  doi = {10.1038/s41467-024-45392-y},
  url = {https://doi.org/10.1038/s41467-024-45392-y}
}

@article{Xray_1T_Albert1997,
  title = {X-ray crystal-structure refinement of the nearly commensurate phase of {$1T\ensuremath{-}{\mathrm{TaS}}_{2}$} in $(3+2)$-dimensional superspace},
  author = {Spijkerman, Albert and de Boer, Jan L. and Meetsma, Auke and Wiegers, Gerrit A. and van Smaalen, Sander},
  journal = {Phys. Rev. B},
  volume = {56},
  issue = {21},
  pages = {13757--13767},
  numpages = {0},
  year = {1997},
  month = {Dec},
  publisher = {American Physical Society},
  doi = {10.1103/PhysRevB.56.13757},
  url = {https://link.aps.org/doi/10.1103/PhysRevB.56.13757}
}

@article{FE_NatRev2023,
  title = {Ferroelectric order in {van der Waals} layered materials},
  author = {Zhang, Dawei and Schoenherr, Peggy and Sharma, Pankaj and Seidel, Jan},
  journal = {Nature Reviews Materials},
  volume = {8},
  issue = {1},
  pages = {25--40},
  year = {2023},
  doi = {10.1038/s41578-022-00484-3},
  url = {https://doi.org/10.1038/s41578-022-00484-3}
}

@article{Bobkov_PRB2024,
  title = {Spin supercurrent in superconductor/ferromagnet {van der Waals} heterostructures},
  author = {Bobkov, G. A. and Bobkov, A. M. and Bobkova, I. V.},
  journal = {Phys. Rev. B},
  volume = {110},
  issue = {10},
  pages = {104506},
  numpages = {11},
  year = {2024},
  month = {Sep},
  publisher = {American Physical Society},
  doi = {10.1103/PhysRevB.110.104506},
  url = {https://link.aps.org/doi/10.1103/PhysRevB.110.104506}
}

@article{Cai_NatCom2021,
  title = {Evidence for anisotropic spin-triplet Andreev reflection at the {2D van der Waals} ferromagnet/superconductor interface},
  author = {Cai, Ranran and Yao, Yunyan and Lv, Peng and Ma, Yang and Xing, Wenyu and Li, Boning and Ji, Yuan and Zhou, Huibin and Shen, Chenghao and Jia, Shuang and Xie, X. C. and Žutić, Igor and Sun, Qing-Feng and Han, Wei},
  journal = {Nature Communications},
  volume = {12},
  issue = {1},
  pages = {6725},
  year = {2021},
  month = {Nov},
  doi = {10.1038/s41467-021-27041-w},
  url = {https://doi.org/10.1038/s41467-021-27041-w}
}

@article{Amit_NatCom2021,
  title = {The observation of $\pi$-shifts in the Little-Parks effect in {4Hb-TaS$_2$}},
  author = {Almoalem, Avior and Feldman, Irena and Mangel, Ilay and Shlafman, Michael and Yaish, Yuval E. and Fischer, Mark H. and Moshe, Michael and Ruhman, Jonathan and Kanigel, Amit},
  journal = {Nature Communications},
  volume = {15},
  issue = {1},
  pages = {4623},
  year = {2024},
  month = {May},
  doi = {10.1038/s41467-024-48260-x},
  url = {https://doi.org/10.1038/s41467-024-48260-x}
}

@article{You_PRB2021,
  title = {Two-dimensional topological superconductivity candidate in a {van der Waals} layered material},
  author = {You, Jing-Yang and Gu, Bo and Su, Gang and Feng, Yuan Ping},
  journal = {Phys. Rev. B},
  volume = {103},
  issue = {10},
  pages = {104503},
  numpages = {6},
  year = {2021},
  month = {Mar},
  publisher = {American Physical Society},
  doi = {10.1103/PhysRevB.103.104503},
  url = {https://link.aps.org/doi/10.1103/PhysRevB.103.104503}
}

@article{Xi_NatPhy2016,
  title = {Ising pairing in superconducting {$\mathrm{NbSe_2}$} atomic layers},
  author = {Xi, Xiaoxiang and Wang, Zefang and Zhao, Weiwei and Park, Ju-Hyun and Law, Kam Tuen and Berger, Helmuth and Forró, László and Shan, Jie and Mak, Kin Fai},
  journal = {Nature Physics},
  volume = {12},
  issue = {2},
  pages = {139--143},
  year = {2016},
  month = {Feb},
  doi = {10.1038/nphys3538},
  url = {https://doi.org/10.1038/nphys3538}
}

@article{PhysRevMaterials.2.094001,
  title = {Two-dimensional superconductivity and topological states in {$\mathrm{PdTe_2}$} thin films},
  author = {Liu, Chong and Lian, Chao-Sheng and Liao, Meng-Han and Wang, Yang and Zhong, Yong and Ding, Cui and Li, Wei and Song, Can-Li and He, Ke and Ma, Xu-Cun and Duan, Wenhui and Zhang, Ding and Xu, Yong and Wang, Lili and Xue, Qi-Kun},
  journal = {Phys. Rev. Mater.},
  volume = {2},
  issue = {9},
  pages = {094001},
  numpages = {8},
  year = {2018},
  month = {Sep},
  publisher = {American Physical Society},
  doi = {10.1103/PhysRevMaterials.2.094001},
  url = {https://link.aps.org/doi/10.1103/PhysRevMaterials.2.094001}
}

@article{PhysRevB.96.220506,
  title = {Type-I superconductivity in the Dirac semimetal {$\mathrm{PdTe_2}$}},
  author = {Leng, H. and Paulsen, C. and Huang, Y. K. and de Visser, A.},
  journal = {Phys. Rev. B},
  volume = {96},
  issue = {22},
  pages = {220506},
  numpages = {5},
  year = {2017},
  month = {Dec},
  publisher = {American Physical Society},
  doi = {10.1103/PhysRevB.96.220506},
  url = {https://link.aps.org/doi/10.1103/PhysRevB.96.220506}
}

@article{PhysRevB.96.041201,
  title = {Nontrivial Berry phase and type-II Dirac transport in the layered material $\mathrm{PdT}{\mathrm{e}}_{2}$},
  author = {Fei, Fucong and Bo, Xiangyan and Wang, Rui and Wu, Bin and Jiang, Juan and Fu, Dongzhi and Gao, Ming and Zheng, Hao and Chen, Yulin and Wang, Xuefeng and Bu, Haijun and Song, Fengqi and Wan, Xiangang and Wang, Baigeng and Wang, Guanghou},
  journal = {Phys. Rev. B},
  volume = {96},
  issue = {4},
  pages = {041201},
  numpages = {7},
  year = {2017},
  month = {Jul},
  publisher = {American Physical Society},
  doi = {10.1103/PhysRevB.96.041201},
  url = {https://link.aps.org/doi/10.1103/PhysRevB.96.041201}
}

@article{Menard_NatCom2017,
  title = {Two-dimensional topological superconductivity in {Pb/Co/Si}(111)},
  author = {Ménard, Gerbold C. and Guissart, Sébastien and Brun, Christophe and Leriche, Raphaël T. and Trif, Mircea and Debontridder, François and Demaille, Dominique and Roditchev, Dimitri and Simon, Pascal and Cren, Tristan},
  journal = {Nature Communications},
  volume = {8},
  issue = {1},
  pages = {2040},
  year = {2017},
  month = {Dec},
  doi = {10.1038/s41467-017-02192-x},
  url = {https://doi.org/10.1038/s41467-017-02192-x}
}

@article{Liao_NatPhys2018,
  title = {Superconductivity in few-layer stanene},
  author = {Liao, Menghan and Zang, Yunyi and Guan, Zhaoyong and Li, Haiwei and Gong, Yan and Zhu, Kejing and Hu, Xiao-Peng and Zhang, Ding and Xu, Yong and Wang, Ya-Yu and He, Ke and Ma, Xu-Cun and Zhang, Shou-Cheng and Xue, Qi-Kun},
  journal = {Nature Physics},
  volume = {14},
  issue = {4},
  pages = {344--348},
  year = {2018},
  month = {Apr},
  doi = {10.1038/s41567-017-0031-6},
  url = {https://doi.org/10.1038/s41567-017-0031-6}
}

@article{Saito_NatPhys2016,
  title = {Superconductivity protected by spin–valley locking in ion-gated {$\mathrm{MoS_2}$}},
  author = {Saito, Yu and Nakamura, Yasuharu and Bahramy, Mohammad Saeed and Kohama, Yoshimitsu and Ye, Jianting and Kasahara, Yuichi and Nakagawa, Yuji and Onga, Masaru and Tokunaga, Masashi and Nojima, Tsutomu and Yanase, Youichi and Iwasa, Yoshihiro},
  journal = {Nature Physics},
  volume = {12},
  issue = {2},
  pages = {144--149},
  year = {2016},
  month = {Feb},
  doi = {10.1038/nphys3580},
  url = {https://doi.org/10.1038/nphys3580}
}

@article{Holleis_2025,
   title={Nematicity and orbital depairing in superconducting Bernal bilayer graphene},
   volume={21},
   ISSN={1745-2481},
   url={http://dx.doi.org/10.1038/s41567-024-02776-7},
   DOI={10.1038/s41567-024-02776-7},
   number={3},
   journal={Nature Physics},
   publisher={Springer Science and Business Media LLC},
   author={Holleis, Ludwig and Patterson, Caitlin L. and Zhang, Yiran and Vituri, Yaar and Yoo, Heun Mo and Zhou, Haoxin and Taniguchi, Takashi and Watanabe, Kenji and Berg, Erez and Nadj-Perge, Stevan and Young, Andrea F.},
   year={2025},
   month=feb, pages={444–450} 
}

@article{Fumega_NanLet2023,
 author= {Fumega, Adolfo Otero and Diego, Josu and Pardo, Victor and Blanco-Canosa, Santiago and Errea, Ion},
 title={Anharmonicity Reveals the Tunability of the Charge Density Wave Orders in Monolayer {VSe$_2$}},
 journal = {Nano Letters}, 
 year=  {2023},
 month= {Mar},
 volume ={23},
 issue= {5},
 pages= {1794--1800} ,
 doi= {10.1021/acs.nanolett.2c04584},
 url = {https://doi.org/10.1021/acs.nanolett.2c04584}
}

@article{Coelho_CDW2019,
 author= {Coelho, Paula Mariel and Nguyen Cong, Kien and Bonilla, Manuel and Kolekar, Sadhu and Phan, Manh-Huong and Avila, José and Asensio, Maria C. and Oleynik, Ivan I. and Batzill, Matthias},
 title={Charge Density Wave State Suppresses Ferromagnetic Ordering in {VSe$_2$} Monolayers},
 journal = {The Journal of Physical Chemistry C}, 
 year=  {2019},
 month= {Jun},
 volume ={123},
 issue= {22},
 pages= {14089--14096} ,
 doi= {10.1021/acs.jpcc.9b04281},
 url = {https://doi.org/10.1021/acs.jpcc.9b04281}
}

@article{Amir_npjQM2025,
 author= {Dalal, Amir and Ruhman, Jonathan and Venderbos, Jörn W. F.},
 title={Flat band physics in the charge-density wave state of {1T-TaS$_2$} and {1T-TaSe$_2$}},
 journal = {npj Quantum Materials}, 
 year=  {2025},
 month= {Mar},
 volume ={10},
 issue= {1},
 pages= {31} ,
 doi= {10.1038/s41535-025-00747-6},
 url = {https://doi.org/10.1038/s41535-025-00747-6}
}

@article{Li_SciAdv2019,
 author= {Li, J. and Li, Y. and Du, S. and Wang, Z. and Gu, B. L. and Zhang, S. C. and He, K. and Duan, W. and Xu, Y.},
 title={Intrinsic magnetic topological insulators in {van der Waals} layered {MnBi$_2$}{Te$_4$}-family materials},
 journal = {Science advances}, 
 year=  {2019},
 month= {Jun},
 volume ={5},
 issue= {6},
 pages= {5885} ,
 doi= {10.1126/sciadv.aaw5685},
 url = {https://doi.org/10.1126/sciadv.aaw5685}
}

@article{PhysRevB.107.134118,
  title   = {Structural and Electronic Phase Transition in the {van der Waals} Crystal {$\mathrm{HfS_2}$} under High Pressure},
  author  = {Zhong, Wei and Deng, Wen and Hong, Fang and Yue, Binbin},
  journal = {Phys. Rev. B},
  volume  = {107},
  number  = {13},
  pages   = {134118},
  year    = {2023},
  month   = apr,
  doi     = {10.1103/PhysRevB.107.134118},
  url     = {https://doi.org/10.1103/PhysRevB.107.134118}
}

@article{PhysRevB.105.104105,
  title = {Structural phase transitions and Raman identifications of the layered {van der Waals} magnet {${\mathrm{CrI}}_{2}$}},
  author = {Zhang, Shuqing and Tang, Fawei and Song, Xiaoyan and Zhang, Xinping},
  journal = {Phys. Rev. B},
  volume = {105},
  issue = {10},
  pages = {104105},
  numpages = {5},
  year = {2022},
  month = {Mar},
  publisher = {American Physical Society},
  doi = {10.1103/PhysRevB.105.104105},
  url = {https://link.aps.org/doi/10.1103/PhysRevB.105.104105}
}

@Article{ma16010454,
AUTHOR = {Lis, Olga and Kozlenko, Denis and Kichanov, Sergey and Lukin, Evgenii and Zel, Ivan and Savenko, Boris},
TITLE = {Structural, Magnetic and Vibrational Properties of {van der Waals} Ferromagnet {CrBr$_3$} at High Pressure},
JOURNAL = {Materials},
VOLUME = {16},
YEAR = {2023},
NUMBER = {1},
ARTICLE-NUMBER = {454},
URL = {https://www.mdpi.com/1996-1944/16/1/454},
PubMedID = {36614792},
ISSN = {1996-1944},
DOI = {10.3390/ma16010454}
}

@article{Wang_ACSNano2022,
  title = {The Magnetic Genome of Two-Dimensional {van der Waals} Materials},
  author = {Wang, Qing Hua and Bedoya-Pinto, Amilcar and Blei, Mark and Dismukes, Avalon H. and Hamo, Assaf and Jenkins, Sarah and Koperski, Maciej and Liu, Yu and Sun, Qi-Chao and Telford, Evan J. and Kim, Hyun Ho and Augustin, Mathias and Vool, Uri and Yin, Jia-Xin and Li, Lu Hua and Falin, Alexey and Dean, Cory R. and Casanova, Fèlix and Evans, Richard F. L. and Chshiev, Mairbek and Mishchenko, Artem and Petrovic, Cedomir and He, Rui and Zhao, Liuyan and Tsen, Adam W. and Gerardot, Brian D. and Brotons-Gisbert, Mauro and Guguchia, Zurab and Roy, Xavier and Tongay, Sefaattin and Wang, Ziwei and Hasan, M. Zahid and Wrachtrup, Joerg and Yacoby, Amir and Fert, Albert and Parkin, Stuart and Novoselov, Kostya S. and Dai, Pengcheng and Balicas, Luis and Santos, Elton J. G.},
  journal = {ACS Nano},
  volume = {16},
  issue = {5},
  pages = {6960--7079},
  year = {2022},
  month = {May},
  publisher = {American Chemical Society},
  doi = {10.1021/acsnano.1c09150},
  url = {https://doi.org/10.1021/acsnano.1c09150}
}

@article{Liu_SciRep2020,
  title = {Experimental observations and density functional simulations on the structural transition behavior of a two-dimensional transition-metal dichalcogenide},
  author = {Liu, W. and Duan, Z. and Zhang, C. and Hu, X. X. and Cao, J. B. and Liu, L. J. . . and Lin, L.},
  journal = {Scientific Reports},
  volume = {10},
  issue = {1},
  pages = {18255},
  year = {2020},
  month = {Oct},
  doi = {10.1038/s41598-020-75240-0},
  url = {https://doi.org/10.1038/s41598-020-75240-0}
}

@article{PhysRevB.102.060103,
  title = {{${T}_{d}$} to {$1{T}^{\ensuremath{'}}$} structural phase transition in the {${\mathrm{WTe}}_{2}$} Weyl semimetal},
  author = {Tao, Yu and Schneeloch, John A. and Aczel, Adam A. and Louca, Despina},
  journal = {Phys. Rev. B},
  volume = {102},
  issue = {6},
  pages = {060103},
  numpages = {5},
  year = {2020},
  month = {Aug},
  publisher = {American Physical Society},
  doi = {10.1103/PhysRevB.102.060103},
  url = {https://link.aps.org/doi/10.1103/PhysRevB.102.060103}
}

@article{Cheon_ACSNano2021,
  title   = {Structural Phase Transition and Interlayer Coupling in Few-Layer {1T$^\prime$} and {$T_d$-{MoTe}$_2$}},
  author  = {Cheon, Yeryun and Lim, Soo Yeon and Kim, Kangwon and Cheong, Hyeonsik},
  journal = {ACS Nano},
  volume  = {15},
  number  = {2},
  pages   = {2962--2970},
  year    = {2021},
  month   = jan,
  doi     = {10.1021/acsnano.0c09162},
  url     = {https://doi.org/10.1021/acsnano.0c09162}
}

@article{Duerloo_NatCom2014,
  title = {Structural phase transitions in two-dimensional Mo- and W-dichalcogenide monolayers},
  author = {Duerloo, Karel-Alexander N. and Li, Yao and Reed, Evan J.},
  journal = {Nature Communications},
  volume = {5},
  issue = {1},
  pages = {4214},
  year = {2014},
  month = {Jul},
  doi = {10.1038/ncomms5214},
  url = {https://doi.org/10.1038/ncomms5214}
}

@article{RevModPhys.96.021003,
  title = {Colloquium: Spin-orbit effects in superconducting hybrid structures},
  author = {Amundsen, Morten and Linder, Jacob and Robinson, Jason W. A. and \ifmmode \check{Z}\else \v{Z}\fi{}uti\ifmmode \acute{c}\else \'{c}\fi{}, Igor and Banerjee, Niladri},
  journal = {Rev. Mod. Phys.},
  volume = {96},
  issue = {2},
  pages = {021003},
  numpages = {34},
  year = {2024},
  month = {May},
  publisher = {American Physical Society},
  doi = {10.1103/RevModPhys.96.021003},
  url = {https://link.aps.org/doi/10.1103/RevModPhys.96.021003}
}

@article{MoonChubukov2010QuantumCritical,
  author       = {Eun‐Gook Moon and Andrey V. Chubukov},
  title        = {Quantum-critical pairing with varying exponents},
  journal      = {Journal of Low Temperature Physics},
  volume       = {161},
  number       = {1-2},
  pages        = {263--281},
  year         = {2010},
  month        = {October},
  doi          = {10.1007/s10909-010-0199-y},
  url          = {https://doi.org/10.1007/s10909-010-0199-y}
}

@article{ChubukovAbanovWangWu2020InterplaySuperconductivity,
  author       = {Andrey V. Chubukov and Artem Abanov and Yuxuan Wang and Yi-Ming Wu},
  title        = {The interplay between superconductivity and non-Fermi liquid at a quantum-critical point in a metal},
  journal      = {Annals of Physics},
  volume       = {417},
  pages        = {168142},
  year         = {2020},
  month        = {September},
  doi          = {10.1016/j.aop.2020.168142},
  url          = {https://doi.org/10.1016/j.aop.2020.168142}
}

@article{KleinKoziiRuhmanFernandes2023QuantumFerroelectricMetals,
  author       = {Avraham Klein and Vladyslav Kozii and Jonathan Ruhman and Rafael M. Fernandes},
  title        = {Theory of criticality for quantum ferroelectric metals},
  journal      = {Physical Review B},
  volume       = {107},
  number       = {16},
  pages        = {165110},
  year         = {2023},
  month        = {April},
  doi          = {10.1103/PhysRevB.107.165110},
  url          = {https://doi.org/10.1103/PhysRevB.107.165110}
}

@article{ChubukovSchmalian2005MasslessBoson,
  author       = {Andrey V. Chubukov and J{\"o}rg Schmalian},
  title        = {Superconductivity due to massless boson exchange in the strong-coupling limit},
  journal      = {Physical Review B},
  volume       = {72},
  number       = {17},
  pages        = {174520},
  year         = {2005},
  month        = {November},
  doi          = {10.1103/PhysRevB.72.174520},
  url          = {https://doi.org/10.1103/PhysRevB.72.174520}
}

@article{Anderson1959DirtySuperconductors,
  author       = {P. W. Anderson},
  title        = {Theory of dirty superconductors},
  journal      = {Journal of Physics and Chemistry of Solids},
  volume       = {11},
  number       = {1-2},
  pages        = {26--30},
  year         = {1959},
  month        = {January},
  doi          = {10.1016/0022-3697(59)90036-8},
  url          = {https://doi.org/10.1016/0022-3697(59)90036-8}
}

@article{MillisSachdevVarma1988Inelastic,
  author       = {A. J. Millis and Subir Sachdev and Chandra M. Varma},
  title        = {Inelastic scattering and pair breaking in anisotropic and isotropic superconductors},
  journal      = {Physical Review B},
  volume       = {37},
  number       = {9},
  pages        = {4975--4986},
  year         = {1988},
  month        = {March},
  doi          = {10.1103/PhysRevB.37.4975},
  url          = {https://doi.org/10.1103/PhysRevB.37.4975}
}

@article{AbanovChubukovNorman2008GapAnisotropy,
  author       = {Artem Abanov and Andrey V. Chubukov and Michael R. Norman},
  title        = {Gap anisotropy and universal pairing scale in a spin-fluctuation model of cuprate superconductors},
  journal      = {Physical Review B},
  volume       = {78},
  number       = {22},
  pages        = {220507},
  year         = {2008},
  month        = {December},
  doi          = {10.1103/PhysRevB.78.220507},
  url          = {https://doi.org/10.1103/PhysRevB.78.220507}
}

@article{PhysRevLett.30.1108,
  title = {Anisotropic Superfluidity in $^{3}\mathrm{He}$: A Possible Interpretation of Its Stability as a Spin-Fluctuation Effect},
  author = {Anderson, P. W. and Brinkman, W. F.},
  journal = {Phys. Rev. Lett.},
  volume = {30},
  issue = {22},
  pages = {1108--1111},
  numpages = {0},
  year = {1973},
  month = {May},
  publisher = {American Physical Society},
  doi = {10.1103/PhysRevLett.30.1108},
  url = {https://link.aps.org/doi/10.1103/PhysRevLett.30.1108}
}

@article{kozii2015odd,
  title={Odd-Parity Superconductivity in the Vicinity of Inversion Symmetry Breaking in Spin-Orbit-Coupled Systems.},
  author={Kozii, V and Fu, L},
  journal={Physical Review Letters},
  volume={115},
  number={20},
  pages={207002--207002},
  year={2015}
}

@article{venditti2025spin,
  title={Spin-dependent anisotropic electron-phonon coupling in KTaO $ \_3$},
  author={Venditti, Giulia and Macheda, Francesco and Barone, Paolo and Lorenzana, Jos{\'e} and Gastiasoro, Maria N},
  journal={arXiv preprint arXiv:2510.25655},
  year={2025}
}

@article{nawwar2025large,
  title={Large phonon-drag thermopower polarity reversal in Ba-doped KTaO3},
  author={Nawwar, Mohamed and Poage, Samuel and Schwaigert, Tobias and Gastiasoro, Maria N and Salmani-Rezaie, Salva and Schlom, Darrell G and Ahadi, Kaveh and Wooten, Brandi L and Heremans, Joseph P},
  journal={arXiv preprint arXiv:2508.00313},
  year={2025}
}

@article{StanevTes2010,
  author  = {Stanev, Valentin and Te{\v{s}}anovi{\'c}, Zlatko},
  title   = {Three-band superconductivity and the order parameter that breaks time-reversal symmetry},
  journal = {Physical Review B},
  volume  = {81},
  number  = {13},
  pages   = {134522},
  year    = {2010},
  doi     = {10.1103/PhysRevB.81.134522}
}

@article{FernandesMaitiWolfleChubukov2013,
  author  = {Fernandes, Rafael M. and Maiti, Saurabh and W{\"o}lfle, Peter and Chubukov, Andrey V.},
  title   = {How many quantum phase transitions exist inside the superconducting dome of the iron pnictides?},
  journal = {Physical Review Letters},
  volume  = {111},
  number  = {5},
  pages   = {057001},
  year    = {2013},
  doi     = {10.1103/PhysRevLett.111.057001}
}

@article{GrinenkoNaturePhys2020,
  author  = {Grinenko, Vadim and Sarkar, Ritu and Das, Debarchan and Leiner, Jesche and Thomale, Ronny and others},
  title   = {Superconductivity with broken time-reversal symmetry inside a superconducting {$s$}-wave state},
  journal = {Nature Physics},
  volume  = {16},
  pages   = {789--794},
  year    = {2020},
  doi     = {10.1038/s41567-020-0886-9}
}

@article{GrinenkoArxiv2018,
  author  = {Grinenko, Vadim and Sarkar, Ritu and Das, Debarchan and others},
  title   = {Emerging superconductivity with broken time-reversal symmetry inside a superconducting {$s$}-wave state},
  journal = {arXiv e-prints},
  eprint  = {1809.03610},
  archivePrefix = {arXiv},
  primaryClass  = {cond-mat.supr-con},
  year    = {2018}
}

@article{GrinenkoPRB2017,
  author  = {Grinenko, Vadim and Sarkar, Ritu and Khasanov, Rustem and others},
  title   = {Superconductivity with broken time-reversal symmetry in ion-irradiated {Ba$_{1-x}$K$_x$Fe$_2$As$_2$}},
  journal = {Physical Review B},
  volume  = {95},
  number  = {21},
  pages   = {214511},
  year    = {2017},
  doi     = {10.1103/PhysRevB.95.214511}
}

@article{nayak2021evidence,
  title={Evidence of topological boundary modes with topological nodal-point superconductivity},
  author={Nayak, Abhay Kumar and Steinbok, Aviram and Roet, Yotam and Koo, Jahyun and Margalit, Gilad and Feldman, Irena and Almoalem, Avior and Kanigel, Amit and Fiete, Gregory A and Yan, Binghai and others},
  journal={Nature physics},
  volume={17},
  number={12},
  pages={1413--1419},
  year={2021},
  publisher={Nature Publishing Group UK London}
}

@article{Chubukov2003first,
  title = {First-Order Superconducting Transition near a Ferromagnetic Quantum Critical Point},
  author = {Chubukov, Andrey V. and Finkel'stein, Alexander M. and Haslinger, Robert and Morr, Dirk K.},
  journal = {Phys. Rev. Lett.},
  volume = {90},
  issue = {7},
  pages = {077002},
  numpages = {4},
  year = {2003},
  month = {Feb},
  publisher = {American Physical Society},
  doi = {10.1103/PhysRevLett.90.077002},
  url = {https://link.aps.org/doi/10.1103/PhysRevLett.90.077002}
}

@article{gastiasoro2022theory,
  title={Theory of superconductivity mediated by Rashba coupling in incipient ferroelectrics},
  author={Gastiasoro, Maria N and Temperini, Maria Eleonora and Barone, Paolo and Lorenzana, Jose},
  journal={Physical Review B},
  volume={105},
  number={22},
  pages={224503},
  year={2022}
}

@article{Saha2025Strong,
  title   = {Strong coupling theory of superconductivity and ferroelectric quantum criticality in metallic {SrTiO\textsubscript{3}}},
  author  = {Saha, Sudip Kumar and Gastiasoro, M. N. and Ruhman, Jonathan and Klein, Avraham},
  journal = {npj Quantum Materials},
  year    = {2025},
  volume  = {10},
  pages   = {82},
  doi     = {10.1038/s41535-025-00798-9},
  url     = {https://doi.org/10.1038/s41535-025-00798-9}
}

@article{Misfit_ChemMat2022,
  title   = {Nanotubes from the Misfit Layered Compound {$\mathrm{(SmS)_{1.19}TaS_2}$}: Atomic Structure, Charge Transfer, and Electrical Properties},
  author  = {Sreedhara, M. B. and Bukvišová, Kristýna and Khadiev, Azat and Citterberg, Daniel and Cohen, Hagai and Balema, Viktor and K. Pathak, Arjun and Novikov, Dmitri and Leitus, Gregory and Kaplan-Ashiri, Ifat and Kolíbal, Miroslav and Enyashin, Andrey N. and Houben, Lothar and Tenne, Reshef},
  journal = {Chemistry of Materials},
  year    = {2022},
  volume  = {34},
  number   = {4},
  pages   = {1838--1853},
  doi     = {10.1021/acs.chemmater.1c04106},
  url     = {https://doi.org/10.1021/acs.chemmater.1c04106}
}

@article{WIEGERS1995152,
title = {Charge transfer between layers in misfit layer compounds},
journal = {Journal of Alloys and Compounds},
volume = {219},
number = {1},
pages = {152-156},
year = {1995},
note = {Eleventh international conference on solid compounds of transition elements},
issn = {0925-8388},
doi = {https://doi.org/10.1016/0925-8388(94)05004-X},
url = {https://www.sciencedirect.com/science/article/pii/092583889405004X},
author = {G.A. Wiegers}
}

@article{Itahashi2025,
  title   = {Misfit layered superconductor {$\mathrm{(PbSe)_{1.14}(NbSe_2)_3}$} with possible layer-selective {FFLO} state},
  author  = {Itahashi, Yuki M. and Nohara, Yamato and Chazono, Michiya and Matsuoka, Hideki and Arioka, Koichiro and Nomoto, Tetsuya and Kohama, Yoshimitsu and Yanase, Youichi and Iwasa, Yoshihiro and Kobayashi, Kaya},
  journal = {Nature Communications},
  year    = {2025},
  volume  = {16},
  number   = {1},
  doi     = {10.1038/s41467-025-62297-6},
  url     = {https://doi.org/10.1038/s41467-025-62297-6}
}

@article{Jindal2023,
  title   = {Coupled ferroelectricity and superconductivity in bilayer {Td-MoTe\textsubscript{2}}},
  author  = {Jindal, Apoorv and Saha, Amartyajyoti and Li, Zizhong and Taniguchi, Takashi and Watanabe, Kenji and Hone, James C. and Birol, Turan and Fernandes, Rafael M. and Dean, Cory R. and Pasupathy, Abhay N. and Rhodes, Daniel A.},
  journal = {Nature},
  year    = {2023},
  volume  = {613},
  number   = {7942},
  pages   =  {48--52},
  doi     = {10.1038/s41586-022-05521-3},
  url     = {https://doi.org/10.1038/s41586-022-05521-3}
}

@article{Zhang_Nat2023,
author= {Zhang Y and Polski R and Thomson A and Lantagne-Hurtubise E and Lewandowski C and Zhou H and Watanabe K and Taniguchi T and Alicea J and Nadj-Perge S},
title={Enhanced superconductivity in spin-orbit proximitized bilayer graphene},
journal = {Nature},
year=  {2023},
month= {Jan},
volume ={613},
issue= {7943},
 pages= {268-273}
}

@article{Klein2018,
  title = {Superconductivity near a nematic quantum critical point: Interplay between hot and lukewarm regions},
  volume = {98},
  ISSN = {2469-9969},
  url = {http://dx.doi.org/10.1103/PhysRevB.98.220501},
  DOI = {10.1103/physrevb.98.220501},
  number = {22},
  journal = {Physical Review B},
  publisher = {American Physical Society (APS)},
  author = {Klein,  Avraham and Chubukov,  Andrey},
  year = {2018},
  month = dec 
}

@article{Klein2019,
  title = {Multiple intertwined pairing states and temperature-sensitive gap anisotropy for superconductivity at a nematic quantum-critical point},
  volume = {4},
  ISSN = {2397-4648},
  url = {http://dx.doi.org/10.1038/s41535-019-0192-x},
  DOI = {10.1038/s41535-019-0192-x},
  number = {1},
  journal = {npj Quantum Materials},
  publisher = {Springer Science and Business Media LLC},
  author = {Klein,  Avraham and Wu,  Yi-Ming and Chubukov,  Andrey V.},
  year = {2019},
  month = nov 
}

@article{
RibakChiralSC_SciAdv2020,
author = {A. Ribak  and R. Majlin Skiff  and M. Mograbi  and P. K. Rout  and M. H. Fischer  and J. Ruhman  and K. Chashka  and Y. Dagan  and A. Kanigel },
title = {Chiral superconductivity in the alternate stacking compound {4Hb-TaS$_2$}},
journal = {Science Advances},
volume = {6},
number = {13},
pages = {eaax9480},
year = {2020},
doi = {10.1126/sciadv.aax9480},
URL = {https://www.science.org/doi/abs/10.1126/sciadv.aax9480},
}

@article{
LukeTRSSr2RuO4_nat1998,
author = {Luke, G. M. and Fudamoto, Y. and Kojima, K. M. and Larkin, M. I. and Merrin, J. and Nachumi, B. and Uemura, Y. J. and Maeno, Y. and Mao, Z. Q. and Mori, Y. and Nakamura, H. and Sigrist, M.},
title = {Time-reversal symmetry-breaking superconductivity in {Sr\textsubscript{2}RuO\textsubscript{4}}},
journal = {Nature},
volume = {394},
number = {6693},
pages = {558--561},
year = {1998},
doi = {10.1038/29038},
URL = {https://doi.org/10.1038/29038}
}

@article{XiaPolarKerrSr2RuO4_PRL2006,
  title = {High Resolution Polar Kerr Effect Measurements of {${\mathrm{Sr}}_{2}{\mathrm{RuO}}_{4}$}: Evidence for Broken Time-Reversal Symmetry in the Superconducting State},
  author = {Xia, Jing and Maeno, Yoshiteru and Beyersdorf, Peter T. and Fejer, M. M. and Kapitulnik, Aharon},
  journal = {Phys. Rev. Lett.},
  volume = {97},
  issue = {16},
  pages = {167002},
  numpages = {4},
  year = {2006},
  month = {Oct},
  publisher = {American Physical Society},
  doi = {10.1103/PhysRevLett.97.167002},
  url = {https://link.aps.org/doi/10.1103/PhysRevLett.97.167002}
}

@article{LevensonPolarKerr_PRL2018,
  title = {Polar Kerr Effect from Time-Reversal Symmetry Breaking in the Heavy-Fermion Superconductor {${\mathrm{PrOs}}_{4}{\mathrm{Sb}}_{12}$}},
  author = {Levenson-Falk, E. M. and Schemm, E. R. and Aoki, Y. and Maple, M. B. and Kapitulnik, A.},
  journal = {Phys. Rev. Lett.},
  volume = {120},
  issue = {18},
  pages = {187004},
  numpages = {6},
  year = {2018},
  month = {May},
  publisher = {American Physical Society},
  doi = {10.1103/PhysRevLett.120.187004},
  url = {https://link.aps.org/doi/10.1103/PhysRevLett.120.187004}
}

@article{Kapitulnik_2009,
doi = {10.1088/1367-2630/11/5/055060},
url = {https://doi.org/10.1088/1367-2630/11/5/055060},
year = {2009},
month = {may},
publisher = {},
volume = {11},
number = {5},
pages = {055060},
author = {Kapitulnik, Aharon and Xia, Jing and Schemm, Elizabeth and Palevski, Alexander},
title = {Polar Kerr effect as probe for time-reversal symmetry breaking in unconventional superconductors},
journal = {New Journal of Physics}
}

@article{RSOC_BordoloiJAP2024,
    author = {Bordoloi, Arjyama and Garcia-Castro, A. C. and Romestan, Zachary and Romero, Aldo H. and Singh, Sobhit},
    title = {Promises and technological prospects of two-dimensional {Rashba} materials},
    journal = {Journal of Applied Physics},
    volume = {135},
    number = {22},
    pages = {220901},
    year = {2024},
    month = {06}
}

@article{PhysRevB.106.L121114,
  title = {Anisotropic resistivity and superconducting instability in ferroelectric metals},
  author = {Zyuzin, Vladimir A. and Zyuzin, Alexander A.},
  journal = {Phys. Rev. B},
  volume = {106},
  issue = {12},
  pages = {L121114},
  numpages = {6},
  year = {2022},
  month = {Sep},
  publisher = {American Physical Society},
  doi = {10.1103/PhysRevB.106.L121114},
  url = {https://link.aps.org/doi/10.1103/PhysRevB.106.L121114}
}

@article{Cohen1967,
  title = {Superconductive Pairing Across Electron Barriers},
  author = {Cohen, Morrel H. and Douglass, D. H.},
  journal = {Phys. Rev. Lett.},
  volume = {19},
  issue = {3},
  pages = {118--121},
  numpages = {0},
  year = {1967},
  month = {Jul},
  publisher = {American Physical Society},
  doi = {10.1103/PhysRevLett.19.118},
  url = {https://link.aps.org/doi/10.1103/PhysRevLett.19.118}
}

@article{Kopasov2024,
  title = {Unconventional superconductivity and paramagnetic Meissner response triggered by nonlocal pairing interaction in proximitized heterostructures},
  author = {Kopasov, A. A. and Mel'nikov, A. S.},
  journal = {Phys. Rev. B},
  volume = {110},
  issue = {9},
  pages = {094503},
  numpages = {14},
  year = {2024},
  month = {Sep},
  publisher = {American Physical Society},
  doi = {10.1103/PhysRevB.110.094503},
  url = {https://link.aps.org/doi/10.1103/PhysRevB.110.094503}
}

@article{Chaudhary2024,
  title = {Superconductivity from Domain Wall Fluctuations in Sliding Ferroelectrics},
  author = {Chaudhary, Gaurav and Martin, Ivar},
  journal = {Phys. Rev. Lett.},
  volume = {133},
  issue = {24},
  pages = {246001},
  numpages = {6},
  year = {2024},
  month = {Dec},
  publisher = {American Physical Society},
  doi = {10.1103/PhysRevLett.133.246001},
  url = {https://link.aps.org/doi/10.1103/PhysRevLett.133.246001}
}

@article{PhysRevLett.130.176801,
  title = {Sliding Phase Transition in Ferroelectric {van der Waals} Bilayers},
  author = {Tang, Ping and Bauer, Gerrit E. W.},
  journal = {Phys. Rev. Lett.},
  volume = {130},
  issue = {17},
  pages = {176801},
  numpages = {7},
  year = {2023},
  month = {Apr},
  publisher = {American Physical Society},
  doi = {10.1103/PhysRevLett.130.176801},
  url = {https://link.aps.org/doi/10.1103/PhysRevLett.130.176801}
}

@Article{Fukuda2020,
  author   = {Fukuda, Takumi and Makino, Kotaro and Saito, Yuta and Fons, Paul and Kolobov, Alexander V. and Ueno, Keiji and Hase, Muneaki},
  journal  = {Applied Physics Letters},
  title    = {Ultrafast dynamics of the low frequency shear phonon in {1T'-MoTe\textsubscript{2}}},
  year     = {2020},
  issn     = {0003-6951},
  month    = {03},
  number   = {9},
  pages    = {093103},
  volume   = {116}
}

@Article{Rano2020,
  author   = {B. Rahman Rano and Ishtiaque M. Syed and S.H. Naqib},
  journal  = {Journal of Alloys and Compounds},
  title    = {Ab initio approach to the elastic, electronic, and optical properties of {MoTe\textsubscript{2}} topological Weyl semimetal},
  year     = {2020},
  issn     = {0925-8388},
  pages    = {154522},
  volume   = {829},
  doi      = {https://doi.org/10.1016/j.jallcom.2020.154522},
  url      = {https://www.sciencedirect.com/science/article/pii/S0925838820308859},
}

@Article{Puotinen1961,
  author  = {Puotinen, D. and Newnham, R. E.},
  journal = {Acta Crystallographica},
  title   = {{The crystal structure of MoTe${\sb 2}$}},
  year    = {1961},
  month   = {Jun},
  number  = {6},
  pages   = {691--692},
  volume  = {14},
  doi     = {10.1107/S0365110X61002084},
  url     = {https://doi.org/10.1107/S0365110X61002084},
}

@Article{Yang2018,
  author    = {Yang, Qing and Wu, Menghao and Li, Ju},
  journal   = {J. Phys. Chem. Lett.},
  title     = {Origin of Two-Dimensional Vertical Ferroelectricity in {WTe2} Bilayer and Multilayer},
  year      = {2018},
  volume    = {9},
  number    = {24},
  pages     = {7160--7164},
  doi       = {10.1021/acs.jpclett.8b03654},
  publisher = {American Chemical Society},
}

@Article{Palle2024,
  title = {Unconventional superconductivity from electronic dipole fluctuations},
  author = {Palle, Grgur and Schmalian, J\"org},
  journal = {Phys. Rev. B},
  volume = {110},
  issue = {10},
  pages = {104516},
  numpages = {26},
  year = {2024},
  month = {Sep},
  publisher = {American Physical Society},
  doi = {10.1103/PhysRevB.110.104516},
  url = {https://link.aps.org/doi/10.1103/PhysRevB.110.104516}
}

@Article{Yuan2019RoomTemperature,
  title        = {Room-temperature ferroelectricity in {MoTe}$_{2}$ down to the atomic monolayer limit},
  author       = {Yuan, Shuoguo and Luo, Xin and Chan, Hung Lit and Xiao, Chengcheng and Dai, Yawei and Xie, Maohai and Hao, Jianhua},
  journal      = {Nature Communications},
  volume       = {10},
  pages        = {1775},
  year         = {2019},
  doi          = {10.1038/s41467-019-09669-x},
  url          = {https://www.nature.com/articles/s41467-019-09669-x}
}

@article{
JeongNodalSC,
author = {Jeong Min Park  and Shuwen Sun  and Kenji Watanabe  and Takashi Taniguchi  and Pablo Jarillo-Herrero },
title = {Experimental evidence for nodal superconducting gap in moiré graphene},
journal = {Science},
volume = {0},
number = {0},
pages = {eadv8376},
year = {},
doi = {10.1126/science.adv8376},
URL = {https://www.science.org/doi/abs/10.1126/science.adv8376},
eprint = {https://www.science.org/doi/pdf/10.1126/science.adv8376}
}

@article{Fay1980,
  author       = {D. Fay and J. Appel},
  title        = {Coexistence of $p$-state superconductivity and itinerant ferromagnetism},
  journal      = {Physical Review B},
  volume       = {22},
  number       = {7},
  pages        = {3173--3182},
  year         = {1980},
  month        = {Oct},
  doi          = {10.1103/PhysRevB.22.3173},
  url          = {https://doi.org/10.1103/PhysRevB.22.3173}
}

@article{Santi2001,
  author       = {G. Santi and S. B. Dugdale and T. Jarlborg},
  title        = {Longitudinal Spin Fluctuations and Superconductivity in Ferromagnetic {ZrZn$_{2}$} from Ab Initio Calculations},
  journal      = {Physical Review Letters},
  volume       = {87},
  number       = {24},
  pages        = {247004},
  year         = {2001},
  doi          = {10.1103/PhysRevLett.87.247004},
  url          = {https://doi.org/10.1103/PhysRevLett.87.247004}
}

@article{Chubukov2004,
  author       = {Andrey V. Chubukov and Catherine P\'epin and Jer\^ome Rech},
  title        = {Instability of the Quantum-Critical Point of Itinerant Ferromagnets},
  journal      = {Physical Review Letters},
  volume       = {92},
  number       = {14},
  pages        = {147003},
  year         = {2004},
  doi          = {10.1103/PhysRevLett.92.147003},
  url          = {https://doi.org/10.1103/PhysRevLett.92.147003}
}

@article{Fu2015,
  author       = {Liang Fu},
  title        = {Parity-Breaking Phases of Spin-Orbit-Coupled Metals with {Gyrotropic}, {Ferroelectric}, and {Multipolar Orders}},
  journal      = {Physical Review Letters},
  volume       = {115},
  number       = {2},
  pages        = {026401},
  year         = {2015},
  doi          = {10.1103/PhysRevLett.115.026401},
  url          = {https://doi.org/10.1103/PhysRevLett.115.026401}
}

@misc{Guo2025,
      title={Layer-selective Cooper pairing in an alternately stacked transition metal dichalcogenide}, 
      author={Haojie Guo and Sandra Sajan and Irián Sánchez-Ramírez and Tarushi Agarwal and Alejandro Blanco Peces and Chandan Patra and Maia G. Vergniory and Rafael M. Fernandes and Ravi Prakash Singh and Fernando de Juan and Maria N. Gastiasoro and Miguel M. Ugeda},
      year={2025},
      eprint={2507.15647},
      archivePrefix={arXiv},
      primaryClass={cond-mat.supr-con},
      url={https://arxiv.org/abs/2507.15647}, 
}

@Article{Naumis2023,
  author    = {Gerardo G Naumis and Saúl A Herrera and Shiva P Poudel and Hiro Nakamura and Salvador Barraza-Lopez},
  journal   = {Reports on Progress in Physics},
  title     = {Mechanical, electronic, optical, piezoelectric and ferroic properties of strained graphene and other strained monolayers and multilayers: an update},
  year      = {2023},
  number    = {1},
  pages     = {016502},
  volume    = {87}
}

@Article{Bian2024,
  author   = {Renji Bian and Ri He and Er Pan and Zefen Li and Guiming Cao and Peng Meng and Jiangang Chen and Qing Liu and Zhicheng Zhong and Wenwu Li and Fucai Liu},
  journal  = {Science},
  title    = {Developing fatigue-resistant ferroelectrics using interlayer sliding switching},
  year     = {2024},
  number   = {6704},
  pages    = {57--62},
  volume   = {385}
}

\end{document}